\documentclass[preprint]{aastex}
\usepackage{mkfig}

\begin{document}

\title{\bf Stellar Ly$\alpha$ Emission Lines in the {\em Hubble Space
  Telescope} Archive:  Intrinsic Line Fluxes and Absorption
  from the Heliosphere and Astrospheres\altaffilmark{1}}

\author{Brian E. Wood\altaffilmark{2}, Seth Redfield\altaffilmark{3},
  Jeffrey L. Linsky\altaffilmark{2},
  Hans-Reinhard M\"{u}ller\altaffilmark{4,5},
  Gary P. Zank\altaffilmark{5}}

\altaffiltext{1}{Based on observations with the NASA/ESA Hubble Space
  Telescope, obtained at the Space Telescope Science Institute, which is
  operated by the Association of Universities for Research in Astronomy,
  Inc., under NASA contract NAS5-26555.}
\altaffiltext{2}{JILA, University of Colorado and NIST, Boulder, CO
  80309-0440; woodb@origins.colorado.edu, jlinsky@jila.colorado.edu.}
\altaffiltext{3}{Harlan J.\ Smith Postdoctoral Fellow, McDonald Observatory,
  University of Texas, Austin, TX 78712-0259; sredfield@astro.as.utexas.edu.}
\altaffiltext{4}{Department of Physics and Astronomy, Dartmouth College,
  6127 Wilder Lab, Hanover, NH 03755-3528; Hans.Mueller@Dartmouth.edu.}
\altaffiltext{5}{Institute of Geophysics and Planetary Physics,
   University of California at Riverside, 1432 Geology, Riverside, CA 92521;
   zank@ucrac1.ucr.edu.}

\begin{abstract}

     We search the {\em Hubble Space Telescope} (HST) archive for
previously unanalyzed observations of stellar H~I Ly$\alpha$ emission
lines, our primary purpose being to look for new detections of
Ly$\alpha$ absorption from the outer heliosphere, and to also search for
analogous absorption from the astrospheres surrounding the observed stars.
The astrospheric absorption is of particular interest because it can be
used to study solar-like stellar winds that are otherwise
undetectable.  We find and analyze 33 HST Ly$\alpha$ spectra in the
archive.  All the spectra were taken with the E140M grating of the Space
Telescope Imaging Spectrograph (STIS) instrument on board HST.  The
HST/STIS spectra yield 4 new detections of heliospheric absorption (70~Oph,
$\xi$~Boo, 61~Vir, and HD~165185) and 7 new detections of astrospheric
absorption (EV~Lac, 70~Oph, $\xi$~Boo, 61~Vir, $\delta$~Eri, HD~128987, and
DK~UMa), doubling the previous number of heliospheric and
astrospheric detections.  When combined with previous results, 10 of 17
lines of sight within 10~pc yield detections of astrospheric absorption.
This high detection fraction implies that most of the ISM within 10~pc must
be at least partially neutral, since the presence of H~I within the ISM
surrounding the observed star is necessary for an astrospheric
detection.  In contrast, the detection percentage is only 9.7\% (3 out of
31) for stars beyond 10~pc.  Our Ly$\alpha$ analyses provide measurements
of ISM H~I and D~I column densities for all 33 lines of sight, and we
discuss some implications of these results.  Finally, we measure
chromospheric Ly$\alpha$ fluxes from the observed stars.
We use these fluxes to determine how Ly$\alpha$ flux
correlates with coronal X-ray and chromospheric Mg~II emission, and we
also study how Ly$\alpha$ emission depends on stellar rotation.

\end{abstract}

\keywords{circumstellar matter ---  ISM: structure --- stars:
  chromospheres --- stars: winds, outflows --- ultraviolet:
  ISM --- ultraviolet: stars}

\section{INTRODUCTION}

     High resolution spectra of H~I Ly$\alpha$ lines from the {\em Hubble
Space Telescope} (HST) have proven to be useful for many purposes.
Hydrogen and deuterium atoms in the interstellar medium produce
absorption features in stellar Ly$\alpha$ spectra that can be analyzed to
yield information about the ISM.  Of particular interest
are measurements of the local deuterium-to-hydrogen (D/H) ratio, which has
been a focal point for these analyses in the past due to its relevance
for cosmology and Galactic chemical evolution
\citep[e.g.,][]{jll95a,ard97,jll98a,avm98,bew04b}.

     In addition to the ISM absorption, some of the Ly$\alpha$ spectra of
nearby stars also show absorption from the outer heliosphere and/or
absorption from the analogous ``astrospheres'' surrounding the observed
stars \citep{jll96,bew96a,kgg97}.
The region of the outer heliosphere that is probed by the
heliospheric absorption is not observable by any other means.  Astrospheres
and the solar-like stellar winds that are implied by their presence are also
otherwise completely undetectable.  Thus, the heliospheric and astrospheric
diagnostics provided by the HST Ly$\alpha$ absorption spectra are truly
unique.  \citet{bew04a} provides a complete review of past results concerning
the analysis of the heliospheric and astrospheric absorption.

     The most important quantitative results from the astrospheric Ly$\alpha$
absorption analyses are the first measurements of mass loss rates for
solar-like stars, although they could also in principle be used as ISM
diagnostics \citep{pcf93}.  Collectively, these measurements
suggest that mass loss increases with magnetic activity and decreases with
stellar age, suggesting that the solar wind was stronger in the past
\citep{bew02b}.  In addition to its obvious importance for solar/stellar
astronomy, the implied stronger wind of the young Sun could have
important ramifications for our understanding of the evolution of planetary
atmospheres in our solar system.  However, the inferred mass-loss/activity
and mass-loss/age relations for cool main sequence stars are based on only
six astrospheric detections.  Many more detections are necessary to confirm
and refine these results.

     Motivated primarily by the need to increase the number of astrospheric
detections, we have searched the HST archive for additional Ly$\alpha$
spectra of nearby cool stars.  We find 33 new data sets to analyze.  In
addition to identifying lines of sight with detectable astrospheric
absorption, we also analyze these data for other purposes:  (1) To search for
heliospheric Ly$\alpha$ absorption, (2) To measure ISM H~I column densities,
and (3) To measure chromospheric Ly$\alpha$ fluxes corrected for ISM,
heliospheric, and astrospheric absorption.  By combining our results with
previous measurements, we ultimately provide here a complete list of
Ly$\alpha$ measurements based on high resolution HST spectra of nearby
cool stars, and we discuss some implications of these results.

\section{THE DATABASE OF HST Ly$\alpha$ SPECTRA}

     In the top section of Table~1, we list 29 lines of sight with HST
Ly$\alpha$ spectra that have been analyzed and published.  (See the
references listed in the last column of the table.)  Except for 36~Oph~A,
all these spectra were taken by the Goddard High Resolution Spectrograph
(GHRS) instrument.  The 36~Oph~A spectrum was taken by the Space Telescope
Imaging Spectrograph (STIS), which replaced the GHRS on HST in 1997.

     The eighth column of Table~1 lists the ISM H~I column densities
(in cm$^{-2}$) measured towards the observed stars.  The ninth column
indicates whether high resolution HST spectra of the interstellar
Mg~II h \& k absorption lines (at 2803~\AA\ and 2796~\AA, respectively)
are available to provide information on the velocity structure of the ISM.
When available, this information is always taken into account in the
analysis of the broad, saturated ISM Ly$\alpha$ absorption, which
increases confidence in the results of the complex Ly$\alpha$ analysis.
Note that only the highest resolution gratings of the GHRS and STIS
instruments, with $R\equiv \lambda/\Delta\lambda \approx 100,000$, can
resolve the narrow Mg~II absorption lines, so column~8 of Table~1 does
not acknowledge the presence of low ($R\approx 1000$) or moderate
($R\approx 25,000$) resolution spectra of Mg~II in the HST archive.

     The tenth and eleventh columns of Table~1 indicate which Ly$\alpha$
spectra show the presence of heliospheric absorption and astrospheric
absorption, in addition to the ISM absorption.  Marginal detections are
indicated by question marks.  Whether a detection is
marginal or not is based on our own determination, after inspecting and
working with the data ourselves.  \citet{ml02} and \citet{avm02} find
evidence for weak heliospheric
absorption towards the similar Capella and G191-B2B lines of sight, but
these claims rely on subtle statistical arguments rather than clearly
visible excess absorption, so we do not consider these to be detections
for our purposes here.

     For eight of the stars in Table~1, the tenth and eleventh columns are
left blank because the Ly$\alpha$ spectra are moderate resolution GHRS
spectra, which generally lack sufficient spectral resolution to allow a
clear detection of heliospheric or astrospheric absorption.  The exception
is Sirius, which shows detectable heliospheric absorption in its moderate
resolution GHRS spectrum \citep{vvi99b}.  This detection owes its success
in large part to the extremely low ISM H~I column density for that line of
sight, which minimizes the ISM absorption and makes the heliospheric
absorption easier to detect.  Nearly all of the stars listed in Table~1
are cool stars, but there are a few exceptions:  Sirius, HZ~43, G191-B2B,
Feige~24, and GD~246.  These stars will not have solar-like astrospheres
around them since they do not have solar-like winds.  For this reason,
column 11 is left blank for these stars.

     Our goal here is to find additional Ly$\alpha$ spectra lurking in
the HST archive to provide new measurements of ISM absorption, and
hopefully new detections of heliospheric and astrospheric absorption.
We confine our attention to lines of sight shorter than 100~pc.  Longer
lines of sight will almost certainly have ISM H~I column densities
too high and ISM Ly$\alpha$ absorption too broad to hope to detect any
heliospheric or astrospheric absorption.  We also restrict ourselves
to observations of cool stars of spectral type F and later, which will
presumably have coronal winds analogous to that of the Sun and will
therefore potentially have detectable astrospheres analogous to the
Sun's heliosphere.

     Finally, the Ly$\alpha$ spectra must have sufficient spectral
resolution to allow for a reasonably confident identification of
heliospheric and astrospheric absorption, if present.  Ideally, this
means high resolution spectra with either the GHRS Ech-A grating or the
STIS E140H grating, which are both capable of fully resolving the H~I and
D~I Ly$\alpha$ absorption line profiles.  Unfortunately, all the Ech-A
and E140H data that fit our criteria have already been analyzed and are
listed in the top section of Table~1.  As mentioned above, the
$R\approx 25,000$ moderate resolution GHRS spectra do not generally have
sufficient spectral resolution for our purposes.  In order to clearly
identify the presence of heliospheric and astrospheric absorption, which
is always highly blended with the ISM absorption, it is necessary to use
the information provided by the ISM D~I absorption profile to constrain
the ISM H~I absorption, so resolving D~I is crucial (see \S4.2).

     Although moderate resolution GHRS spectra do not meet our
resolution requirements, moderate resolution STIS spectra with the E140M
grating {\em are} acceptable.  The STIS E140M spectra have $R\approx 45,000$,
which is sufficient to mostly resolve the D~I absorption profile.
Furthermore, STIS E140M spectra cover a wide wavelength range of
1150--1730~\AA, meaning that there are many E140M spectra in the HST
archive taken to observe different emission lines within
E140M's broad spectral range that will nevertheless also include the
Ly$\alpha$ line at 1216~\AA\ in which we are interested.

     We have searched the HST archive for previously unanalyzed STIS
E140M spectra of cool stars within 100~pc, and in the bottom portion of
Table~1 we list the 33 lines of sight observed by HST that meet these
criteria.  All observations except those of $\xi$~Boo~A and 61~Vir were
performed through the $0.2^{\prime\prime}\times 0.2^{\prime\prime}$ aperture.
The $\xi$~Boo~A and 61~Vir data were obtained through the narrower
$0.2^{\prime\prime}\times 0.06^{\prime\prime}$ aperture.
The data were reduced using the STIS team's CALSTIS software
package written in IDL \citep{dl99}.  The processing includes wavelength
calibration using spectra of the calibration lamp taken during the
observations, and includes correction for scattered light.  We generally
coadd all available E140M spectra in the archive, with two exceptions.
For AD Leo, there are separate data sets on 2000 March 10 and
2002 June 1.  We choose not to include the latter because the geocoronal
Ly$\alpha$ emission is difficult to remove in those data.
(See \S3 for more discussion about the geocoronal emission and its
removal.)  For V471~Tau, there are three different large data sets.  For
the sake of simplicity, we choose to reduce only the observations taken
on 2000 August 24--27.

\section{THE GEOCORONAL Ly$\alpha$ EMISSION}

     Figure~1 shows the final processed spectra of all 33 stars, focusing
on the Ly$\alpha$ region.  Each spectrum shows a stellar emission line upon
which is superimposed a very broad H~I Ly$\alpha$ absorption feature
centered near the rest wavelength of 1215.670~\AA.  The spectra also show
narrow D~I Ly$\alpha$ absorption about $-0.33$~\AA\ from the center of the
H~I absorption, except in cases where the H~I absorption is broad enough
to obscure the D~I absorption (see \S4.2).

     Finally, the spectra are contaminated by H~I Ly$\alpha$ emission from
the Earth's geocorona, which is shaded in Figure~1.  These geocoronal lines
are narrow, and most are conveniently contained within the saturated core
of the ISM H~I absorption, allowing them to be easily removed.  In such
cases, we remove the geocoronal lines by fitting Gaussians to the lines and
then subtracting the Gaussians from the data.  Note that the HD~209458
spectrum is the sum of spectra taken at two different times when the
geocoronal emission was at different wavelengths, so two Gaussians are
fitted to the data in this case.

     In some cases the geocoronal emission is blended with the sides of
the H~I absorption profile, making the geocoronal subtraction significantly
harder and more uncertain (e.g., HD~73350).  In such instances,
we still use Gaussian fitting to estimate the absorption, but we only fit
the side of the emission close to the absorption core and we force the
fitted Gaussian to be at the expected wavelength, which we know to
within about 2 km~s$^{-1}$ based on the accuracy of the wavelength
calibration (see below).  Nevertheless, the uncertainty in the geocoronal
subtraction for these blended cases could be a significant source of
systematic error in the analysis of the H~I absorption lines.  In all
future figures, Ly$\alpha$ spectra will be shown with the geocoronal
emission removed.

     Although the geocoronal emission is an annoyance when it is blended
with the sides of the absorption profile, its presence is actually
beneficial when it is fully within the absorption core, because it can be
used as a secondary wavelength calibrator.  The data reduction places the
spectra in a heliocentric rest frame, which means that the geocoronal
emission is centered at the projected velocity of the Earth towards the
star at the time of observation.  This quantity is known accurately and
can be compared with the measured center of the emission, in order to test
the wavelength scale.  In Figure~2, we plot the velocity discrepancies of
the geocoronal emission from their expected locations
($\Delta v \equiv v_{\rm obs}-v_{\rm geo}$) for all the newly analyzed
stars.  These are all STIS E140M spectra, so this represents an excellent
test of the STIS E140M wavelength calibration.

     The weighted mean and standard deviation of the data points in
Figure~2 is $\Delta v=-1.20\pm 0.90$ km~s$^{-1}$.  This result suggests
that CALSTIS-processed E140M spectra are systematically blueshifted by
1.2 km~s$^{-1}$ from where they should be, at least in the Ly$\alpha$
spectral region.  The standard CALSTIS data reduction computes a global
wavelength offset for the entire STIS spectrum from the wavelength
calibration images.  Perhaps the Ly$\alpha$ spectral region, which is
near the edge of the E140M grating's spectral coverage, is systematically
shifted from the average offset value.  There are also a few individual
spectra that show significant discrepancies of up to 4 km~s$^{-1}$ from
the average $\Delta v$ value.  Whatever the cause of the systematic
wavelength shift and the scatter around it, we believe that the geocoronal
lines provide a better velocity scale, so for all spectra in which a
geocoronal emission centroid can be measured, we correct the wavelengths
of our spectra using the $\Delta v$ values in Figure~2.

\section{ANALYZING THE Ly$\alpha$ ABSORPTION LINES}

\subsection{Reconstructing the Stellar H~I Ly$\alpha$ Profile}

     The D~I Ly$\alpha$ absorption lines are relatively easy to analyze
by themselves.  Because they are narrow, a reasonably accurate continuum
can be interpolated over these lines and an absorption profile can then
be fitted to the D~I lines.  \citet{sr04a} have already
performed these measurements for some of the stars in our sample.  Figure~3
shows examples of some of the D~I fits.  The thin solid line shows the
assumed background continuum, the dotted lines show the
individual absorption components of the fit, and the thick solid line shows
the combination of all the components after convolution
with the instrumental line spread function (LSF).  The STIS E140M LSF used
here is from \citet{kcs99}.

     Analyzing the H~I Ly$\alpha$ absorption is much trickier.  Not only
is the visible absorption broad and difficult to interpolate over, but
ISM H~I column densities towards even the nearest stars are high enough
for the H~I absorption profile to have extended damping wings.  Thus,
analyzing the H~I absorption essentially requires first reconstructing the
entire stellar Ly$\alpha$ profile.  This can be done as follows.

     The first step is to use a fit to the D~I absorption to infer what
the central wavelength and Doppler broadening parameter should be for the
H~I line.  The ISM H~I absorption should presumably have the same centroid
velocity as D~I, so $v({\rm H~I})=v({\rm D~I})$, and since past work has
demonstrated that H~I and D~I absorption lines in the local ISM are
dominated by thermal broadening \citep[e.g.,][]{jll96,sr04b}, their
Doppler parameters are related by
$b({\rm H~I})\approx \sqrt{2} b({\rm D~I})$.  With this information, an H~I
opacity profile, $\tau_{\lambda}$ (assumed to be a Voigt profile), can be
computed for various assumed values of the H~I column density,
$N({\rm H~I})$.  For each opacity profile, we can then reconstruct the
wings of the stellar emission line profile by multiplying the data by
$\exp(\tau_{\lambda})$.  (Note that since we are interested in modeling the
damping wings rather than the central part of the absorption, the assumed
value of $b({\rm H~I})$ is actually unimportant.)

     Figure~4 shows the results of this exercise using the $\xi$~Boo
data as an example, with the D~I absorption having been removed.  Stellar
line profiles are reconstructed assuming ISM H~I column densities in the
range $\log N({\rm H~I})=17.6-18.2$ (units cm$^{-2}$).  Only the wings of
the profile outside
the dashed lines are reconstructed using the opacity profiles computed as
described above.  In order to estimate the profile in between the dashed
lines, we either use a polynomial fit to the wings to interolate between
them, or more commonly we use the Mg~II h \& k lines to estimate the shape
of the central portion of the Ly$\alpha$ profile.  The justification for
this is that Ly$\alpha$ and the two Mg~II lines are all highly optically
thick chromospheric lines that have similar profiles in the solar spectrum
\citep{rfd94,pl98}.
Table~1 indicates which HST data sets include high resolution Mg~II spectra.
If high resolution Mg~II spectra are not available for a star, we often
use the Mg~II profile of a similar star.  (The $\xi$~Boo~A Mg~II profiles
are the most commonly used surrogates.)

\subsection{Fitting the ISM Absorption}

     The profile reconstruction technique described in \S4.1 results in a
family of stellar Ly$\alpha$ profiles, as in Figure~4, each of which can
be used as a starting point for an analysis of the H~I absorption.
With each assumed profile, an initial H~I+D~I absorption fit is performed,
which is done using a $\chi^{2}$ minimization technique \citep{prb92}.
All necessary atomic data for the D~I and H~I Ly$\alpha$
lines are taken from \citet{dcm03}.  The absorption is convolved with the
LSF before comparing with the data, as in the D~I fits in Figure~3.  In all
our fits involving H~I absorption, we fit H~I and D~I simultaneously and
constrain the problem by forcing $v({\rm H~I})=v({\rm D~I})$ and
$b({\rm H~I})=\sqrt{2} b({\rm D~I})$.  After the initial fit, the assumed
stellar profile is then altered based on the residuals of the fit, and then
a second, presumably improved absorption line fit is performed.  Additional
iterations can be done to further refine the fit.  For a simple line of
sight with one ISM absorption component and no heliospheric or astrospheric
absorption, the problem is well constrained enough that this iterative
process ultimately ends up driving the fit parameters of the single
absorption component towards the same solution {\em no matter which of the
assumed stellar profiles one actually starts out with.}  However, this is
not always the case in more complicated multi-component lines of sight,
especially when heliospheric or astrospheric absorption is present
\citep[see][]{jll96}.

     Fortunately, it is possible to constrain the analysis even further
by assuming a D/H ratio.  Using UV spectra from HST and the {\em Far
Ultraviolet Spectroscopic Explorer} (FUSE), a consensus has been reached
that the interstellar D/H ratio is invariable within the Local Bubble.
Based on different compilations of measurements,
\citet{jll98a} finds ${\rm D/H}=(1.50\pm 0.10)\times 10^{-5}$, \citet{hwm02}
quotes ${\rm D/H}=(1.52\pm 0.08)\times 10^{-5}$, and \citet{bew04b}
computes ${\rm D/H}=(1.56\pm 0.04)\times 10^{-5}$, with no evidence
for significant variation within 100~pc.

     Since all our targets are located within the Local Bubble, we simply
assume ${\rm D/H}=1.5\times 10^{-5}$, consistent with the values quoted
above.  With the addition of this assumption, all three fit parameters of
the difficult-to-analyze H~I absorption [$v({\rm H~I})$, $b({\rm H~I})$,
and $N({\rm H~I})$] are dependent on the much better constrained D~I fit
parameters [$v({\rm D~I})$, $b({\rm D~I})$, and $N({\rm D~I})$].
The ISM H~I absorption profile is therefore much more tightly constrained,
even when multiple ISM components are included.  This assumption also
means that we do not have to experiment with all the stellar profiles
reconstructed as in Figure~4.  We can instead immediately focus on the
profile that was constructed assuming the $N({\rm H~I})$ value that
the D~I fit suggests should yield ${\rm D/H}=1.5\times 10^{-5}$.  In the
case of $\xi$~Boo~A that value is $\log N({\rm H~I})=17.9$, and this
profile is emphasized in Figure~4.  Not having to consider the other
profiles as possible starting points for the H~I absorption analysis
greatly simplifies what would otherwise be a formidable and time-consuming
problem considering the number of spectra that we wish to analyze.

     In order to provide the reader with an idea of how H~I+D~I absorption
profiles vary under the constraints described above, Figure 5 shows
profiles computed for a range of column densities,
$\log N({\rm H~I})=17.5-18.7$, and for two different Doppler parameters,
$b({\rm H~I})=9.08$ km~s$^{-1}$ and $b({\rm H~I})=12.85$ km~s$^{-1}$.
With only a few exceptions, the range of column densities covers the values
observed for the $d<100~pc$ lines of sight listed in Table~1.  The two
Doppler parameters represent temperatures of $T=5000$~K and
$T=10,000$~K, respectively (in the absence of nonthermal broadening), which
roughly define the range of temperatures that are generally observed in
the local ISM (see \S4.4).

     In Figure 5, the D~I line is clearly narrower for the lower
$b$ value, and at low column densities the Doppler core of the H~I
absorption is also narrower, but the far damping wings of the H~I
absorption are only dependent on the column density.  At the highest column
densities shown, the Doppler core of the H~I absorption has been
obliterated by the broadening absorption from the damping wings, so nearly
all dependence on $b$ is lost.  By $\log N({\rm H~I})=18.7$, the D~I
absorption has become saturated and almost completely blended with H~I.
Thus, the D~I line is no longer very useful for constraining
$N({\rm H~I})$ via the ${\rm D/H}=1.5\times 10^{-5}$ assumption.
However, the width of the base of the H~I absorption has become much
more sensitive to $N({\rm H~I})$, so a reasonably accurate H~I column
density can still be measured in this high column regime.

     The ISM velocity structure along a line of sight is something that
we take into account in our Ly$\alpha$ fits, when this information is
available.  Unfortunately, the thermal width of the D~I line is too
large to suggest the presence of any but the most widely separated
ISM components, and the H~I absorption is naturally far too broad and
opaque for these purposes.  The ISM Mg~II h \& k absorption
lines are the most commonly used lines for studying the velocity structure
of the local ISM, although weaker Fe~II lines observable in near-UV spectra
can be useful as well.  The large atomic weights of the Mg~II and Fe~II
species mean that the thermal broadening of their ISM absorption lines
is much lower than for D~I and H~I.  In fact, these lines are typically
dominated by nonthermal rather than thermal broadening.  Thus, when observed
at the highest spectral resolution accessible to GHRS or STIS, the very
narrow Mg~II and Fe~II lines are ideal for indentifying multiple ISM
velocity components.

     Table~1 indicates which of our lines of sight have high resolution
Mg~II data that can be used to infer the ISM velocity structure.
\citet[][hereafter RL02]{sr02} have already analyzed these and other data
to provide a complete survey of observed ISM Mg~II and Fe~II absorption for
lines of sight within 100~pc.  In cases where multiple ISM components are
seen, we use the results of RL02 to constrain multi-component fits
to our Ly$\alpha$ data.  In a multi-component Ly$\alpha$ fit we force the
velocity separations of the components to be the same as those seen
for Mg~II.  We do not fix the velocities themselves in order to allow the
fit to account for possible differences in the wavelength calibrations of
the Mg~II and Ly$\alpha$ spectra, so we allow one component centroid to
vary and force the other components to have the appropriate shifts from
that component.  Because the ISM components are so highly blended
in the broad H~I and D~I lines, our philosophy is to initially assume
as many constraints as possible.  Thus, initially we force all components
to have the same Doppler parameter in our fits (though we do not force
a particular value), and we also assume that the column density ratios
of the individual components are the same as found for Mg~II.
If a reasonable fit to the data cannot be obtained with these
constraints we try relaxing them, the justification being that warm,
neutral material within the Local Bubble is not entirely homogeneous.
Temperatures and dust depletions within the local ISM vary to some extent,
resulting in variations in Doppler parameters and Mg~II/H~I column density
ratios \citep{np97,sr04b}.  However, all components
are always constrained by the deuterium-hydrogen self-consistency
requirements described above.

     Note that throughout this paper, the ``reasonableness'' of a fit is
ultimately assessed by eye.  The fitting process includes the tweaking of
the assumed stellar emission profile to maximize the quality of a fit (see
above).  Since this tweaking is not automated, the final value of
$\chi^{2}$ is in large part dependent on our ability to alter the
Ly$\alpha$ profile in a reasonable fashion.  This ability is imprecise and
hard to quantify, so the final value of $\chi^{2}$ is of limited
usefulness in assessing whether a fit is acceptable, or whether it is
necessary to relax certain assumptions in order to obtain a better fit.
Thus, we ultimately rely on our own subjective judgment rather than using
a certain $\chi^{2}$ threshold or some other statistical test.

     For 25 of the 33 lines of sight, we are able to adequately fit the
Ly$\alpha$ spectra with only ISM absorption.  The resulting fits are shown
in Figure~6 along with the reconstructed stellar Ly$\alpha$ profiles.
Only the total ISM absorption is shown in Figure~6.  For the
multi-component fits, Figure~3 provides an idea of what the absorption
contributions from the individual components are like.  (Although the
Figure~3 fits are D~I-only fits, the H~I+D~I fits in Figure~6 will not be
very different in how they fit the D~I line due to the constraints
described above that are imposed on these fits.)

     Table~2 lists the parameters of these fits, using the ID numbers
assigned in Table~1 to identify the stars.  The quoted 1$\sigma$
uncertainties indicate only the random errors induced in the fits by the
noise of the spectra.  There is no attempt to include systematic errors
such as those associated with the shape of the assumed stellar line
profile, which surely dominate the uncertainty in the analysis.

     For the H~I column densities, which are of particular interest for
ISM studies, systematic errors include uncertainties in the assumed
D/H value, keeping in mind that for most of our lines of sight the
dominant constraint on $N({\rm H~I})$ is $N({\rm D~I})$.  The uncertainties
in the average Local Bubble D/H values quoted near the beginning of
this section are $\sim 5$\%.  When combined with uncertainties related to
the reconstructed stellar line profile and the ISM velocity structure, we
believe that typical uncertainties in the {\em total} H~I column
densities listed in Table~1 are $\sim 10$\%, or about 0.04 dex in
$\log N({\rm H~I})$.  (See the notes on the $\xi$~Boo analysis in \S4.5
for discussion on how such errors might affect detection of heliospheric
and astrospheric absorption.)  However, in cases where a good D~I profile
is not present in the data, either due to poor S/N or a high column density
that leads to D~I being saturated and blended with H~I (e.g., HD~203244,
HD~106516, HD~128987, HD~28568, HD~28205, HD~28033, HD~209458, and
$\iota$~Cap), systematic uncertainties in the $\log N({\rm H~I})$ values
listed in Table~1 are probably more like $0.1-0.2$~dex.

     It should be noted that in multi-component cases, the total H~I
column density along a line of sight will be known better than the column
densities of the individual components, which are listed in Table~2,
because the components are so highly blended that the component H~I
column densities will be highly dependent on the assumptions made about
the components described above.  Thus, systematic errors for the
$\log N({\rm H~I})$ values listed in Table~2 for the individual
components will be even higher than those estimated above for the total
H~I column densities.  For the multi-component fits, the component
numbers listed in Table~2 are the same numbers as those used by RL02.
The last column of Table~2 indicates which lines of sight are discussed
in more detail in \S4.5.

     There are three different velocities listed in Table~2 that can be
compared with the measured ISM velocity, $v({\rm H~I})$:  $V_{LIC}$,
$V_{G}$, and $v({\rm Mg~II})$.  The $V_{LIC}$
quantity indicates the line-of-sight velocity predicted by the Local
Interstellar Cloud (LIC) vector \citep{rl92,rl95}, which is pointed
towards Galactic coordinates [$l=186.1^{\circ}$, $b=-16.4^{\circ}$] with a
magnitude of 25.7 km~s$^{-1}$.  This agrees well with measurements of
ISM particles moving through the solar system, which suggest a flow
directed at coordinates [$l=183.3^{\circ}$, $b=-15.9^{\circ}$]
at a speed of 26.3 km~s$^{-1}$ \citep{mw93,mw04}.  However,
for lines of sight near the Galactic center direction, the so-called G
cloud vector of \citet{rl92} can be a better predictor of ISM
absorption line velocities, so the $V_{G}$ quantity indicates
the velocities suggested by this vector.  A sky map from \citet{jll00}
provides a crude estimate of the region on the sky where the G cloud
vector is applicable, which suggests that the G cloud vector is most
likely to apply for stars 33, 39, 45, 46, and 52.  It is worth noting
that it is very questionable whether the G cloud is truly distinct from
the LIC and other nearby clouds, or if the apparent different velocity
vector in that direction is merely a product of complex velocity
gradients within local ISM material \citep{pcf02,pcf03a}.  Despite this
ambiguity, the G cloud vector is still useful for our purposes as a
predictor of ISM velocities in directions where the LIC vector clearly
does not work.  Finally, for lines of sight that have Mg~II measurements,
we also list in Table~2 the Mg~II velocities from RL02, $v({\rm Mg~II})$,
although for $\upsilon$~Peg and $\iota$~Cap we choose to quote the Fe~II
velocities instead (see \S4.5).

\subsection{Identifying the Presence of Heliospheric and/or Astrospheric
  Absorption}

     The null assumption in all our Ly$\alpha$ analyses is that there is
no heliospheric or astrospheric absorption present.  We make every effort
to find a reasonable fit to the H~I+D~I Ly$\alpha$ absorption assuming the
presence of only ISM absorption.  However, for 8 of our 33 lines of sight
we conclude that there is simply no way that the D~I and H~I absorption
can be fitted in a self-consistent manner without including additional
H~I absorption beyond that from the ISM.  The additional H~I absorption
component must have a column density too low [$\log N({\rm H~I})<17.0$)]
to produce any absorption in lines from much less abundant species such as
D~I, Mg~II, or Fe~II.  It is this excess H~I absorption that we interpret
as being from the heliosphere or from the astrospheres surrounding the
observed stars.

     In our fits to these data in Figure~7, we add a single absorption
component to the analysis to represent heliospheric and/or astrospheric
absorption.  This component is labeled as ``HS/AS'' in Table~2,
where the final fit parameters are listed for all our Ly$\alpha$ fits.
In Figure~7, we show the absorption provided only by the ISM, the
discrepancy with the data indicating the excess absorption that we
believe must be coming from HS/AS absorption.  Past theoretical work has
shown that the heliosphere only produces excess absorption on the red
side of the Ly$\alpha$ absorption line, while astrospheric absorption
produces excess absorption on the blue side of the line
\citep{kgg97,vvi99b,bew00c}.  The redshift of heliospheric absorption
relative to the ISM is mostly due to the deceleration and deflection of H
atoms as they cross the bow shock, whereas the corresponding effect for
astrospheres results instead in a blueshift since our perspective is from
outside the astrosphere rather than inside.  Thus, based on the fits in
Figure~7, we report and list in Table~1 the following new astrospheric
absorption detections:  EV~Lac, 70~Oph, $\xi$~Boo, 61~Vir, $\delta$~Eri,
HD~128987, and DK~UMa.  However, we consider the 61~Vir and DK~UMa
detections to be marginal (see \S4.5).  The new heliospheric absorption
detections are: 70~Oph, $\xi$~Boo, 61~Vir, and HD~165185.

     Details about the evidence for the presence of the HS/AS absorption
will be provided in \S4.5 for each individual line of sight, but some
general comments can be made here.  The most important one is that it
is impossible to overstate the importance of the constraints on the
ISM H~I absorption provided by the D~I absorption.  In cases where only
heliospheric {\em or} astrospheric absorption is present (EV~Lac,
$\delta$~Eri, HD~165185, HD~128987, and DK~UMa), the HS/AS absorption
produces a significant shift of the H~I absorption centroid away from that
suggested by the D~I absorption (see Fig.~7).  If multiple ISM components
are present, one can try to account for this by allowing the ISM
components to have different Doppler parameters.  Given the saturated
nature of the H~I absorption, increasing $b({\rm H~I})$ for one component
can significantly increase the absorption in this component relative to
the others, whereas this will not be the case for the unsaturated D~I
line, which is far more affected by the relative column densities of the
components than by their relative $b({\rm D~I})$ values.  Multiple ISM
components can therefore in principle induce an apparent shift of the H~I
absorption relative to D~I.  However, because the ISM absorption
components are always so highly blended, and because of the stringent
deuterium-hydrogen self-consistency constraints described in \S4.2, the
relative velocity shift that can be induced in this fashion without
resulting in poor fits to either D~I or H~I is limited.

     A second source of evidence for HS/AS absorption, especially important
in cases where {\em both} heliospheric and astrospheric absorption are
present (70~Oph, $\xi$~Boo, and 61~Vir), is the overall width of
the base of the H~I absorption.  The presence of HS/AS absorption makes
this width larger than can be explained by ISM absorption alone (see
Fig.~7).  In the H~I column density regime of concern here, one can try
to broaden an ISM H~I component either by increasing $b({\rm H~I})$ or
$N({\rm H~I})$.  However the ${\rm D/H}=1.5\times 10^{-5}$ constraint
means that it is not possible to greatly increase $N({\rm H~I})$ without
producing too much D~I absorption, and the
$b({\rm H~I})=\sqrt{2} b({\rm D~I})$ constraint means that increasing
$b({\rm H~I})$ too much will eventually make the D~I absorption too broad.

     In principle, the stellar Ly$\alpha$ profiles can be altered to
account for the excess HS/AS absorption in Figure~7, but this does not
work in practice since the alterations necessary to remove the need for
the HS/AS absorption yield profiles that are completely implausible
\citep{bew96a}.  Using $\delta$~Eri as an example, changing the stellar
profile to fit the data without requiring the existence of the excess
H~I absorption seen in Figure~7 necessitates the assumption of a profile
that decreases from about where it is in Figure~7 at 1215.45~\AA\
($\sim 9\times 10^{-12}$ ergs cm$^{-2}$ s$^{-1}$ \AA$^{-1}$) down to
zero flux at about 1215.57~\AA.  This is not a reasonable
stellar profile.  Reasonable alterations to stellar profiles do not
include the introduction of extreme profile slopes or zero flux within
the emission profile.

     Note that although we list the parameters for the HS/AS components
in Table~2 for completeness, they are of little use in estimating
properties of the heliospheric and/or astrospheric gas.  More sophisticated
analyses of this excess absorption using hydrodynamic modeling of the
heliosphere and astrospheres is required to extract meaningful
quantitative information from the heliospheric and astrospheric
absorption \citep*{bew00c,bew02b}, which will be done in a future paper.
We do not bother using separate heliospheric and astrospheric
absorption components even when both are clearly present, partly because
of the limited usefulness of the fit parameters of these empirical
absorption components, and partly because our goal here is simply to infer
the presence of the HS/AS absorption and illustrate the amount of excess
absorption in Figure~7.

\subsection{The ISM Fit Parameters}

     The fit parameters listed for the ISM absorption components in Table~2
provide useful diagnostics for the local ISM.  The column densities
indicate the average ISM H~I densities along the lines of sight when
divided by the target distances, the velocities provide information on the
ISM flow vector, and the Doppler parameters indicate the temperature of the
gas, which can be estimated by the equation $T=60.6 b({\rm H~I})^{2}$ if
the small contributions of turbulent velocities to the line broadening are
neglected \citep{sr04b}.  However, interpreting the meaning of these
measurements is somewhat problematic unless the velocity structure is known.
The high resolution Mg~II and Fe~II spectra that provide this information
are only available for 12 of our 33 lines of sight.  Since 7 of these 12
sight lines show multiple ISM components, the prevalence of multiple velocity
components is clear.  The presence of unresolved velocity structure can
shift the centroid of the ISM absorption away from the expected $V_{LIC}$
or $V_{G}$ values, and it can also broaden the absorption lines, thereby
artificially increasing the D~I and H~I Doppler parameters.

     Of the 12 lines of sight with Mg~II and/or Fe~II information, all but
one show an absorption component within 3 km~s$^{-1}$ of the expected
$V_{LIC}$ or $V_{G}$ values, the exception being AU~Mic (star \#39).  This
comparison is valid whether the measured velocities considered are
$v({\rm H~I})$ or $v({\rm Mg~II})$, with the exception of $\iota$~Cap, for
which the $v({\rm H~I})$ velocities are poorly constrained and therefore
discrepant (see \S4.5).  Of the 21 lines of sight without the benefit of
the Mg~II or Fe~II data, 10 show line centroids over 3 km~s$^{-1}$ from the
expected $V_{LIC}$/$V_{G}$ values.  The worse agreement for these sight
lines is surely due to unresolved velocity structure, especially
considering that the $b({\rm H~I})$ values for 7 of these 10 lines of sight
are suspiciously high based on previous work.  Previous analyses of high
resolution HST spectra, constrained by knowledge of the ISM velocity
structure, have generally suggested that warm, partially neutral clouds
within the Local Bubble typically have temperatures of $T=5000-10,000$~K,
corresponding to Doppler parameters of $b({\rm H~I})=9.1-12.8$ km~s$^{-1}$
\citep{ard97,np97,bew98,sr04b}.  Thus, Doppler
parameters of $b({\rm H~I})>12.8$ km~s$^{-1}$ are suggestive of possible
unresolved velocity components.  (See the notes on the AD~Leo analysis in
\S4.5 for an example of how unresolved velocity structure can artificially
increase measured Doppler parameters.)

     One issue that should be mentioned is that the $b({\rm H~I})$
values reported in Table~2, which are derived from the H~I+D~I fits, are
almost always $1-2$ km~s$^{-1}$ larger than those that we measure
from the D~I-only fits [assuming
$b({\rm H~I})=\sqrt{2} b({\rm D~I})$], such as the fits shown in Figure~3.
The most likely explanation for this is that the LSF we are assuming is
slightly broader than it should be, meaning that we are overcorrecting for
instrumental broadening.  For the two single-component fits in
Figure~3 ($\xi$~Boo and DK~UMa), the difference between the pre- and
post-convolution fits indicates the extent that the D~I absorption profile
is unresolved based on our assumed LSF.  The D~I line is clearly not
entirely resolved in the E140M data, which means that the convolution
correction is important and that uncertainties in the LSF will lead to
uncertainties in the fit parameters, especially $b({\rm D~I})$.
This emphasizes the importance of high spectral resolution in the Ly$\alpha$
analyses.

     Finally, since we are {\it a priori} assuming a D/H value in our fits,
our analysis cannot refine the value of the Local Bubble D/H ratio, but we
can at least say that our success in fitting the data in Figures~6 and 7
indicates that the Ly$\alpha$ data analyzed here are all consistent with
the assumed value of ${\rm D/H}=1.5\times 10^{-5}$.

\subsection{Notes on Individual Lines of Sight}

     We here provide comments on some of the individual lines of sight
that have been analyzed.  These include not only notes of particular
interest about the lines of sight, but also in some
cases details about how multiple ISM components were constrained and
especially the precise evidence for heliospheric and astrospheric
absorption, when present.  The comments frequently refer to information
listed in Tables 1 and 2.

\begin{description}
\item[AD Leo (\#31):]  This line of sight is notable for having a
  remarkably high ISM H~I column density, $\log N({\rm H~I})=18.47$,
  given its very short distance of 4.69~pc.  The implied average density
  of $n({\rm H~I})=0.204$ cm$^{-3}$ is to our knowledge the highest
  average density yet measured for any line of sight within 100~pc (i.e.,
  within the Local Bubble), being slightly
  above the $n({\rm H~I})=0.199$ cm$^{-3}$ value towards HD~82558 (see
  Table~1).  The large $b({\rm H~I})$ value of the fit
  and the discrepancy between $v({\rm H~I})$ and the expected $V_{LIC}$
  velocity strongly imply the presence of multiple ISM components.
  There are no high resolution Mg~II spectra available to verify this,
  but in Figure~8 we show an alternative fit to the data where
  an arbitrary second ISM component is included.  The second component is
  assumed to be redshifted by 12 km~s$^{-1}$ from the primary component,
  with a column density a factor of 10 lower than that of the primary
  component.  The Doppler parameters of the two components are forced to
  be identical, and in the fit the Doppler parameter of both components
  ends up at $b({\rm H~I})=12.85$ km~s$^{-1}$.  This is a more
  plausible result than the high $b({\rm H~I})=14.05$ km~s$^{-1}$ value
  from the single component fit, illustrating how the presence of
  multiple components can artificially increase measured Doppler
  parameters.  The primary reason that the Figure~6 fit has such a
  poor $\chi^{2}$ value ($\chi^{2}_{\nu}=2.571$) is the very high
  signal-to-noise (S/N) of the data.  The high S/N is a result of the
  unusually long 52~ksec exposure time of the coadded HST spectrum,
  which was taken as part of an extensive flare monitoring
  campaign \citep{slh03}.
\item[EV Lac (\#32):]  Figure~7 shows that there is a lot of excess H~I
  absorption on the blue side of the line, resulting in the centroid of
  the H~I absorption being highly blueshifted with respect to the D~I
  absorption.  This therefore represents a convincing detection of
  astrospheric absorption.
\item[70 Oph (\#33):]  The Mg~II analysis by RL02
  suggests a complex line of sight with 3 ISM components, including a
  component (component 3 in Table~2) that is shifted relative to the primary
  component by an unusually large amount for such a short sight line.
  The primary component (component 1) is a G cloud component.
  The complexity of the ISM structure complicates the Ly$\alpha$ analysis.
  In fitting D~I by itself (see Fig.~3), we find that we cannot fit the
  data well enough while forcing the column density ratios of the 3
  components to be consistent with those of the Mg~II fit, so in
  Figure~3 we have allowed the column density ratios to be different.
  This D~I fit predicts a Doppler parameter for the H~I absorption of
  $b({\rm H~I})=8.4\pm 0.9$ km~s$^{-1}$ [assuming as always
  $b({\rm H~I})=\sqrt{2} b({\rm D~I})$], implying a very low ISM temperature
  of $T=4300\pm 900$~K, consistent with the results of \citet{sr04b}.
  This measurement is not implausible since other
  G cloud lines of sight have also shown particularly low temperatures
  \citep{jll96,bew00b,sr04b}.  In the D~I+H~I fit shown in Figure~7 and
  Figure~9a we force the column density ratios to be the same as those
  derived from the D~I-only fit.  However,
  fits to the data with only the ISM components invariably require
  $b({\rm H~I})\approx 15$ km~s$^{-1}$ in order to produce H~I absorption
  broad enough to fit the data.  This fit obviously produces D~I absorption
  much too broad to fit the D~I line, considering the low
  $b({\rm H~I})=8.4$ km~s$^{-1}$ measurement quoted above from the D~I-only
  fit.  Thus, the data require the addition of an HS/AS component in order
  to explain the overly broad H~I absorption.  The best fit in Figure~7
  shows excess absorption on both sides of the line, implying the presence
  of both heliospheric and astrospheric absorption.  Note that because the
  blueshifted component~3 is not a strong contributor to the D~I absorption,
  it is possible to allow {\em only} that component to have
  $b({\rm H~I})\approx 15$ km~s$^{-1}$ while still fitting D~I reasonably
  well.  In doing so, component 3 becomes broad enough to account for the
  excess blue-side H~I absorption, as shown in Figure~9b, meaning that only
  heliospheric H~I absorption would be inferred from the data.  However,
  $b({\rm H~I})\approx 15$ km~s$^{-1}$ is inconsistent with the width of
  the Mg~II line.  Even in the unlikely case that the component~3 Mg~II
  absorption line is entirely thermally broadened, the Mg~II Doppler
  parameter of $b({\rm Mg~II})=2.76\pm 0.15$ km~s$^{-1}$ measured by
  RL02 suggests an upper limit for the H~I Doppler
  parameter of $b({\rm H~I})<14.3$ km~s$^{-1}$, so we do not consider the
  fit in Figure~9b to be an acceptable interpretation of the data.  Thus,
  we believe the astrospheric detection implied by the fit in Figure~7 is
  solid, as well as the heliospheric detection.
\item[$\xi$ Boo (\#34):]  Excess absorption on both sides of the Ly$\alpha$
  line is indicated by the fit in Figure~7, so this is both a heliospheric
  and an astrospheric detection.  The heliospheric absorption is slightly
  stronger than the astrospheric absorption, inducing a slight redshift of
  the total H~I absorption relative to D~I.  Thus, the evidence for the
  heliospheric absorption is somewhat stronger than the evidence for the
  astrospheric absorption.  Evidence for the astrospheric absorption relies
  on the relative widths of the H~I and D~I lines.  Figure~10 shows a fit to
  the $\xi$~Boo data in which the centroid of the HS/AS absorption component
  is fixed in such a way that it can only account for the red-side excess
  (i.e., the heliospheric absorption).  The fit works well for H~I, but the
  $b({\rm H~I})=12.7\pm 0.2$ km~s$^{-1}$ value of the ISM absorption leads
  to a D~I absorption line that is too broad.  The D~I-only fit in Figure~3
  suggests $b({\rm H~I})=9.6\pm 0.6$ km~s$^{-1}$.  We believe that this
  discrepancy is too large to be explained by the possible LSF problem noted
  in \S4.4, so we cannot consider the fit to D~I in Figure~10 to be
  acceptable.  Thus, we conclude that the astrospheric detection implied by
  the Figure~7 fit is secure.  Since the width of the ISM absorption
  depends in part on $N({\rm H~I})$ as well as $b({\rm H~I})$ (see
  Fig.~5), uncertainties in $N({\rm H~I})$ could potentially lead
  to an erroneous detection of HS/AS absorption.  In \S4.2, we estimate
  uncertainties in $N({\rm H~I})$ to be typically $\sim 10$\% after
  including systematic errors, such as uncertainties in the
  ${\rm D/H}=1.5\times 10^{-5}$ assumption.  In Figure~11, we repeat the
  fit from Figure~7, but we also show how the ISM absorption profile
  changes when the H~I column density is changed by 10\%.  A 10\% increase
  in $N({\rm H~I})$ clearly does not come close to explaining the excess
  H~I absorption that we believe is HS/AS absorption.
\item[61 Vir (\#35):]  The evidence for the heliospheric and astrospheric
  absorption shown in Figure~7 is almost identical to that for $\xi$~Boo,
  so the 61~Vir analysis mirrors the $\xi$~Boo analysis described above.
  However, the S/N of the 61~Vir data is not nearly as high as that of the
  $\xi$~Boo spectrum.  This means that the width of the D~I absorption
  (and therefore the ISM H~I absorption as well) cannot be measured
  with the same precision.  Thus, the evidence for the astrospheric
  absorption is weaker, and in Table~1 we indicate only a marginal
  detection of astrospheric absorption.
\item[$\delta$ Eri (\#37):]  The enormous amount of excess H~I absorption
  on the blue side of the line leads to a large blueshift of H~I relative
  to D~I, amounting to a very convincing detection of astrospheric
  absorption (see Fig.~7).
\item[$\kappa$ Cet (\#38):]  This is a complex, 3 component line of sight
  based on the Mg~II data (RL02).  As is the case for
  70~Oph (see above), we find that the D~I absorption is fitted
  significantly better when the individual components are allowed to have
  column density ratios different from that of Mg~II.  Thus, the fit in
  Figure~3 allows the column densities of the individual components to vary,
  and in our H~I+D~I fits we force the column density ratios to be the same
  as for the D~I-only fit.  However, we are unable to fit the data with the
  usual assumption of equal Doppler parameters for all ISM components.  In
  order to fit the data we allow the most blueshifted component (component 3)
  to have a higher Doppler parameter, which in our best fit is
  $b({\rm H~I})=13.3\pm 0.2$ km~s$^{-1}$.  This is suspiciously high for
  local ISM material, but the $b({\rm H~I})$ value is not
  inconsistent with the $b({\rm Mg~II})=2.48\pm 0.38$ km~s$^{-1}$
  measurement (RL02), unlike what happens in the similar 70~Oph analysis
  (see above).  Thus, we conclude that the fit is plausible and no
  HS/AS absorption component is required to fit the data.
\item[AU Mic (\#39):]  This is the one line of sight for which {\em both}
  H~I and Mg~II velocities imply a significant discrepancy from the
  values predicted by the G and LIC vectors (see Table~2).  The Mg~II
  absorption lines analyzed by RL02 do not show asymmetries that would
  demonstrate the presence of multiple ISM components, but both the
  Mg~II and H~I Doppler parameters, $b({\rm Mg~II})=4.94\pm 0.74$ km~s$^{-1}$
  and $b({\rm H~I})=13.74\pm 0.08$ km~s$^{-1}$, are suspiciously high.
  The Mg~II spectrum analyzed by RL02 is a GHRS Ech-B spectrum taken through
  GHRS's large aperture before the installation of the COSTAR corrective
  optics.  The spectrum therefore has significantly lower resolution than
  is normally achievable with the high resolution gratings of GHRS and STIS.
  Furthermore, it is difficult to estimate the shape of the stellar Mg~II
  profile above the ISM absorption due to the narrowness of the Mg~II
  emission lines, and the location of the ISM absorption on the side of
  the line.  For these reasons, we suspect that there are in fact
  multiple ISM components towards AU~Mic that are not apparent due
  to these difficulties with the Mg~II data.  Finally, it is worth noting
  that a debris disk has recently been discovered around AU~Mic
  \citep{plk04,mcl04}.  Despite being edge-on, there is to our
  knowledge no evidence for any absorption from the disk in the UV spectra
  of AU~Mic from HST, including the Ly$\alpha$ line analyzed here.
  The disk must therefore be gas-poor.
\item[$\zeta$~Dor (\#40):]  For this two-component fit, we force the
  velocity separations and column density ratios of the components to be
  the same as in the Mg~II fit of RL02, but it is necessary to allow
  the Doppler parameters of both components to vary in order to fit the H~I 
  absorption.  Component~1 ends up with a very high Doppler parameter of
  $b({\rm H~I})=14.8\pm 0.1$ km~s$^{-1}$.  However, the Mg~II Doppler
  parameter of $b({\rm Mg~II})=5.76\pm 0.37$ km~s$^{-1}$ reported by RL02
  for this component is also extremely high, so we believe that the high
  $b({\rm H~I})$ value is plausible.
\item[HD 165185 (\#45):]  The H~I absorption shows a small but significant
  redshift relative to the D~I absorption, indicating the presence of
  heliospheric absorption, as shown in Figure~7.
\item[HD 203244 (\#46):]  The $\log N({\rm H~I})=18.82$ column density for
  HD~203244 is high enough that the H~I and D~I absorption lines are
  completely blended (see Fig.~6).  This line of sight and that towards
  HD~82558 have two of the three highest column densities listed in Table~1,
  despite these stars being only about 20~pc away.  Since
  column densities this high are clearly rare within the Local Bubble,
  they are presumably due to relatively small clouds that few sight lines
  pass through.  In order to explain the high columns, these small clouds
  must necessarily have average densities that are significantly higher than
  the $n({\rm H~I})\approx 0.1$~cm$^{-3}$ value that is typical for the LIC
  \citep{jll00}.
\item[HD 106516 (\#50):]  A clear D~I absorption line cannot be discerned
  in the data, partly due to poor S/N and partly due to a high H~I column
  density that leads to D~I and H~I being at least partly blended.  The
  measured H~I column density of $\log N({\rm H~I})=18.57$ is the only fit
  parameter that can be trusted, and this value is probably good to only
  within $\pm 0.2$~dex considering systematic errors.  This is at the upper
  end of the range of $N({\rm H~I})$ uncertainties for our measurements
  quoted in \S4.2.
\item[HD 128987 (\#52):]  Analogous to the EV~Lac and $\delta$~Eri lines of
  sight mentioned above, the HD~128987 data show extensive blue-side excess
  H~I absorption, implying a substantial amount of astrospheric absorption
  (see Fig.~7).  The detection is very convincing despite the rather poor
  S/N of the data.  In the Figure~7 fit, the HS/AS component is forced to
  have a centroid on the blue side of the H~I line to keep it from
  contributing absorption to the red side of the H~I line.  (If the
  resulting fit in Figure~7 had been poor, that would have been evidence
  for heliospheric absorption, in addition to the astrospheric absorption.)
\item[DK UMa (\#53):]  Figure~12 shows the best fit to the data possible
  with only ISM absorption.  The fit shows a bit too much absorption
  on the blue side of the D~I line, and slightly too little absorption
  on the blue side of the H~I line.  In other words, the H~I absorption in
  reality seems to be slightly blueshifted relative to D~I, suggesting the
  presence of astrospheric absorption.  Thus, the best fit in Figure~7
  includes an HS/AS component.  The implied excess absorption is quite weak
  compared with the other detections of astrospheric and heliospheric
  absorption, but the S/N of the data is good enough for the problems with
  the Figure~12 fit to be considered significant.  Furthermore, there are
  Mg~II observations that show only a single ISM component towards DK~UMa,
  which means that the deficiencies of the Figure~12 fit are not due to any
  effects of multiple ISM components.  Thus, we consider this to be an
  astrospheric detection, though we consider it a marginal one, as noted in
  Table~1.  The HS/AS component in the Figure~7 fit contributes a little
  absorption to the red side of the line as well as to the blue side where
  the astrospheric absorption resides, but the red-side contribution is
  too weak for us to consider that to be evidence for heliospheric
  absorption.
\item[V471 Tau (\#58):]  This target is a K2~V+DA eclipsing binary with a
  very short period of 0.51~days.  The UV STIS spectrum as a whole is a
  continuum spectrum from the hot white dwarf star, but the white dwarf
  continuum decreases dramatically near Ly$\alpha$ so that the Ly$\alpha$
  emission line from the K2~V star contributes most of the background for
  the ISM Ly$\alpha$ absorption.  The spectrum shown in Figure~1 is the
  result of a coaddition of spectra taken at different orbital phases, which
  we do in order to provide sufficient S/N for our analysis.  Since the
  ISM absorption is stationary in the individual exposures, the coaddition
  is valid for our purposes despite the fact that the overlying Ly$\alpha$
  profile of the K2~V star moves due to the rapid orbital motion.
  The nonstationary emission results in a reconstructed line
  profile in Figure~6 that looks strange and is in fact unphysical, but an
  average integrated stellar Ly$\alpha$ line flux can still be estimated
  from it (see \S7).  Variable ultraviolet absorption features in
  {\em International Ultraviolet Explorer} (IUE)
  and GHRS spectra have been interpreted as being due to coronal
  mass ejections (CMEs) from the K2~V star, with an implied mass loss rate
  much higher than that of the Sun \citep{djm89,heb01}.
  We see no astrospheric H~I absorption that would provide additional
  evidence for a strong wind from V471~Tau.  This by no means disproves the
  CME interpretation of the variable UV features, given the likely
  possibility that V471~Tau is surrounded by a fully ionized ISM that will
  produce no astrospheric H~I absorption regardless of the stellar
  mass loss rate (see \S6).
\item[HD 209458 (\#59):]  This star has a planetary companion that transits
  in front of the star once every 3.52~days \citep{dc00}.
  By comparing STIS G140M Ly$\alpha$ spectra taken during and outside
  transit times, \citet{avm03} find that the Ly$\alpha$
  profile decreases as much as 15\% at some wavelengths during transit,
  suggesting absorption from an extended, evaporating planetary atmosphere.
  This does not affect our analysis of the ISM absorption in the E140M
  spectrum, however, and we are able to fit the rather noisy D~I and H~I
  absorption lines with a single ISM absorption component.
\item[$\upsilon$ Peg (\#60):]  The ISM Mg~II and Fe~II absorption lines
  show three velocity components for this line of sight (RL02).  Due to the
  saturated nature of the Mg~II lines, we decide that the less opaque Fe~II
  lines provide more precise constraints for our fits in Figures~3 and 6.
  Thus, the $v({\rm Mg~II})$ velocities listed in Table~2
  for $\upsilon$~Peg are in this case the Fe~II velocities.  For the
  Ly$\alpha$ fit in Figure~6, we force the velocity separations and
  column density ratios of the components to be the same as Fe~II, but we
  have to allow the Doppler parameters of
  the individual components to be different to acceptably fit the data.
  The Doppler parameters of the two weaker components (components 1 and 3)
  are suspiciously high (see Table~2), but they are not inconsistent with
  the Fe~II Doppler parameters from RL02.  Considering uncertainties
  induced by the complexity of the line of sight, we conclude that this is
  a reasonable fit.
\item[HD 32008 (\#61):]  This target is a G7~IV-III+DA binary according
  to \citet{gc99}.  As is the case for V471~Tau (see above),
  the white dwarf dominates the UV spectrum everywhere except in the region
  of Ly$\alpha$, where the emission line from the G7~IV-III star provides
  most of the background continuum for the ISM absorption.  There are no
  Mg~II or Fe~II spectra to precisely define the ISM velocity structure
  of this line of sight, but inspection of the D~I absorption
  reveals that this is one case where ISM velocity components are so
  widely separated that evidence for multiple components is apparent even
  in the broad D~I line (see Fig.~3).  Thus, in Figure~3 D~I is fitted
  with two components.  The H~I+D~I fit in Figure~6 is likewise performed
  with two components, where we force them to have the same Doppler
  parameters, but allow their velocities and column densities to both
  vary.
\item[$\iota$ Cap (\#62):]  Three ISM velocity components are suggested
  by the Mg~II and Fe~II absorption lines (RL02).  As is the case for
  $\upsilon$~Peg (see above), the Mg~II lines are saturated, so we decide
  to use the less opaque Fe~II lines to constrain the fit in Figure~6.
  Thus, the $v({\rm Mg~II})$ velocities listed in Table~2
  for $\iota$~Cap are actually Fe~II velocities.  The H~I
  column density for this line of sight, $\log N({\rm H~I})=18.70$, is high
  enough that D~I is nearly saturated and is almost completely blended with
  the H~I absorption.  This blending and the complexity of the line of
  sight make the analysis poorly constrained and very uncertain.  We
  believe that the large H~I Doppler parameters,
  $b({\rm H~I})=15.9$ km~s$^{-1}$, and the
  large discrepancies between the $v({\rm H~I})$ and $v({\rm Mg~II})$
  velocities in Table~2 are a consequence of these difficulties and the
  discrepancies should therefore be regarded with skepticism.
\end{description}

\section{THE HELIOSPHERIC ABSORPTION DETECTIONS}

     The 33 new lines of sight that we have analyzed have resulted in
4 new detections of heliospheric Ly$\alpha$ absorption.  This brings
the total number of detections to 8, if one considers the Alpha/Proxima Cen
detections as a single line of sight.  This is a large enough number that
we can investigate what the lines of sight that produce detectable
absorption have in common.  There are really only two factors that determine
whether a line of sight will have detectable heliospheric Ly$\alpha$
absorption.  One is the ISM H~I column density, since a high ISM column
will mean a broad absorption line that will hide the heliospheric absorption.
The second factor is the orientation of the line of sight through the
heliosphere.  This orientation is most simply described by the angle,
$\theta$, between the line of sight and the upwind direction of the ISM
flow seen by the Sun.  This upwind direction has Galactic coordinates of
$l=6.1^{\circ}$, $b=16.4^{\circ}$ \citep{rl95}.

     In Figure~13, we plot the ISM column densities, $\log N({\rm H~I})$,
versus $\theta$ for all the HST-observed lines of sight listed in Table~1.
Different symbols are used to indicate which lines of sight have detectable
heliospheric absorption and which do not.  The detections are nicely
separated from the nondetections in this parameter space.  As expected, the
lines of sight that yield detections tend to have low $N({\rm H~I})$
values.  It is also clearly easier to detect heliospheric absorption in
upwind directions ($\theta < 90^{\circ}$) than in downwind directions
($\theta > 90^{\circ}$).  Models of the heliosphere can explain this
behavior quite nicely.  It is in upwind directions that the heliospheric
H~I suffers the strongest deceleration at the bow shock
\citep*{vbb93,vbb95,gpz96,bew00c}.  Thus, the
heliospheric absorption in these directions will be shifted away from the
ISM absorption to the greatest extent, making it easier to detect.  The
success of the models in explaining the detection tendencies illustrated
in Figure~13 is additional strong evidence that heliospheric absorption is
indeed the correct interpretation for the red-side excess Ly$\alpha$
absorption.

     For the upwind directions, heliospheric absorption
is only detected when $\log N({\rm H~I}) < 18.2$, but in downwind
directions a detection requires $\log N({\rm H~I}) < 17.8$.  There
simply are not many lines of sight that will have interstellar column
densities of $\log N({\rm H~I}) < 17.8$.  Thus, the downwind detection of
heliospheric absorption towards Sirius may forever remain unique
\citep*{vvi99b}.

     The properties of a line of sight through the heliosphere are
described entirely by the angle $\theta$ only if the heliosphere is
precisely axisymmetric. However, significant deviations from axisymmetry
are possible due to latitudinal solar wind variations, an ISM magnetic
field that is skewed with respect to the ISM flow direction, or unstable
MHD phenomena near the plane of the heliospheric current sheet
\citep{hlp97,rr98,mo03}.  There are now
enough heliospheric absorption detections to conduct at least
a crude investigation into whether there is evidence for any such
asymmetries, and this will be a project for the near future.

\section{THE ASTROSPHERIC ABSORPTION DETECTIONS}

     The 7 new astrospheric detections found here bring the total number
of detections up to 13.  The astrospheric detections are primarily of
interest because the astrospheric absorption provides a diagnostic for
stellar winds that are otherwise completely undetectable.  Mass loss
rates have been measured for all of the older detections, and these
measurements have been used to infer how winds vary with age and activity
for solar-like main sequence stars \citep{bew02b}.  We are currently
measuring mass loss rates using the new astrospheric
detections, but these results will have to wait for a future paper.

     We can use the new detections to investigate what is common among
the lines of sight that have detections, as we did for the
heliospheric detections in \S5, but the situation is more complicated in the
case of astrospheres.  There are many more factors involved in the
detectability of the astrospheric absorption, including the nature of the
ISM surrounding the star, the total integrated ISM H~I column density for
the line of sight, the speed and orientation of the ISM flow vector seen by
the star, and the properties of the stellar wind.  A nondetections of
astrospheric absorption could be due to any number of these factors, so
it is generally impossible to use nondetections of astrospheric
absorption to infer anything about the stellar wind or surrounding ISM,
unlike what can be done with the nondetections of heliospheric absorption
(see \S5).

     In Figure~14, we plot the ISM column density versus stellar
distance for all of the HST-observed lines of sight listed
in Table~1.  The HST data points are placed into three categories:
lines of sight that yield detections of astrospheric absorption, lines of
sight that yield nondetections, and other lines of sight for which a
search for astrospheric absorption is not appropriate due to a hot star
target or low spectral resolution (see \S2).  Finally, for completeness
we also include in Figure~14 five additional data points that are based on
H~I column density measurements from the {\em Extreme Ultraviolet Explorer}
(EUVE) or {\em Copernicus}.  These lines of sight are WD~1634-573
\citep{bew02a}, WD~2211-495 \citep{gh02}, WD~0621-376
\citep{nl02}, $\alpha$~Vir \citep{dgy76}, and
$\alpha$~Cru \citep{dgy76}.

     Considering only those HST lines of sight for which a valid search for
astrospheric absorption can be made, the detection rate is quite high
within 10~pc, with 10 of 17 independent sight lines producing detections
for a detection rate of 58.8\%.  The situation worsens dramatically beyond
10~pc, with only 3 of 31 lines of sight yielding detections, for a
detection rate of only 9.7\%.  Furthermore, 2 of the 3 detections beyond
10~pc are considered only marginal detections (see Table~1).

     There are two reasons for the distance dependence of astrospheric
detectability.  One is simply that longer lines of sight tend to have higher
ISM column densities, as shown by Figure~14.  This means that the ISM
H~I Ly$\alpha$ absorption will be broader, which can hide the astrospheric
absorption.  A second major problem with more distant targets is that there
is a high probability that they will be surrounded by hot, ionized ISM
material with no neutrals, which will therefore inject no H~I into the
astrospheres to produce Ly$\alpha$ absorption.  The Local Bubble is believed
to be mostly filled with this ionized material \citep[e.g.,][]{dms99,rl03}.
There are regions of warm, partially neutral
material embedded within the Local Bubble, and it just so happens that the
Sun is located within one of these regions, explaining why even the nearest
lines of sight have substantial H~I column densities (see Table~1).
However, the further a star is from the Sun, the more likely it will be
located in the more prevalent ionized ISM.  This is evident in Figure~14,
which shows that H~I column densities do not increase much beyond 10~pc, and
average H~I densities therefore decrease.

     Since an astrospheric detection requires the presence of neutral
material in the ISM surrounding the observed star, a detection of
astrospheric absorption represents an {\it in situ} detection of neutral
ISM at the star's location.  Thus, the astrospheric detections can in
principle be used to crudely map out locations of warm, partially neutral
clouds embedded within the Local Bubble.  The 58.8\% detection fraction
within 10~pc suggests that neutral clouds occupy about 58.8\% of space
within 10~pc, although in reality this is really just a lower bound since
nondetections of astrospheres can be for many reasons other than an
ionized surrounding ISM (see above).  It must also be noted that the
observed targets are preferentially nearby rather than being randomly
distributed throughout the volume of space within 10~pc, so the filling
fraction lower limit of 0.588 is biased towards shorter
distances.  Nevertheless, these results clearly suggest that
most of the ISM within 10~pc is warm, partially neutral gas, which is
atypical within the Local Bubble.

     Various observations of ISM H~I atoms flowing through the heliosphere
provide estimates of the H~I density immediately surrounding the Sun.
These estimates are typically $n({\rm H~I})\approx 0.2$ cm$^{-3}$
\citep{eq94,vvi99a,pcf03b}.
Except for the new measurement towards AD~Leo (see \S4.5), average
line-of-sight densities within 10~pc are always well below this value
(see Fig.~14), so crude models of the distribution of gas in the very
local ISM have typically assumed instead $n({\rm H~I})=0.1$ cm$^{-3}$
\citep{jll00,sr00}.  Most of the 10~pc
lines of sight shown in Figure~14 have densities even lower than this,
including most of the astrospheric detections.

     Of particular interest is the $\xi$~Boo line of sight.  The Mg~II
absorption data show only a single velocity component, with a velocity
consistent with the LIC vector (RL02).  Since $\xi$~Boo is an astrospheric
detection, that would suggest that the LIC must extend the entire 6.70~pc
distance towards $\xi$~Boo, but this means that the LIC would have
an average density of only $n({\rm H~I})=0.040$ cm$^{-3}$ for this sight
line, much lower than the $n({\rm H~I})\sim0.2$ cm$^{-3}$ value measured
for the LIC at the Sun's location.  Photoionization models of the LIC
predict that hydrogen ionization
should increase and the H~I density therefore decrease towards the edge of
the cloud due to less shielding from ionizing radiation
\citep[e.g.,][]{jds02}, but not to the extent that one would expect to
see {\em this} large a decrease in an {\em average} line-of-sight density.
There are two alternative explanations for the low average density.  It is
possible that there are in fact two separate clouds along the line of sight
that just happen to have the same projected velocity, in which case a
sizable gap in between the clouds could explain the low average density.
Such a situation must be the case for DK~UMa if its marginal astrospheric
detection is to be believed.  Only a single LIC Mg~II absorption component
is observed towards DK~UMa (RL02), but there is no way that the LIC extends
the 32.4~pc distance to that star.  The second explanation is that the LIC
has a substantial amount of subparsec scale patchiness, with the Sun being
located within a small, high-density patch.  Both of these two
explanations indicate that the ISM velocity structure information provided
by Mg~II and Fe~II absorption data may not be sufficient to infer the true
character of the LIC and other very nearby interstellar material.

     It is worth noting that everything said above about the $\xi$~Boo
sight line also applies to the 36~Oph sight line, except that the G cloud
vector applies towards 36~Oph instead of the LIC vector.  Both $\alpha$~Cen
and 36~Oph are G cloud lines of sight with astrospheric detections and with
Mg~II spectra that show only a single G cloud component.  However, the
implied average H~I densities towards these stars are very different:
$n({\rm H~I})=0.098$ cm$^{-3}$ and $n({\rm H~I})=0.038$ cm$^{-3}$,
respectively.  Thus, the nearby ISM in the direction where the G cloud
vector applies appears to be just as inhomogeneous as the $\xi$~Boo line
of sight, despite the apparent simplicity of the Mg~II velocity structure.

     With only 3 astrospheric detections beyond 10~pc, and with 2 of them
being marginal detections, it is harder to infer general properties of
the ISM beyond 10~pc from these results.  The low detection fraction
is a serious problem for potential observational searches for astrospheric
absorption beyond 10~pc.  In addition to our desire to use astrospheres to
probe the ISM beyond 10~pc, the stellar wind research would also benefit
from more distant detections, since there are many classes of stars that
do not have many representatives within 10~pc.
If the goal is to investigate what astrospheric absorption and stellar
winds are like for giants or young, solar-like G stars, for example,
it is necessary to look beyond 10~pc to find targets.  If astrospheres
could be detected from the ground, an extensive observing program could be
proposed and a low $5-10$\% detection fraction tolerated, but this is not
practical when the only instrument capable of detecting the astrospheres is
the UV spectrograph on board HST.  For the time being, this
point may be moot.  The apparent recent demise of the STIS instrument in
2004 August means that no new high resolution Ly$\alpha$ spectra will be
possible for the foreseeable future.

\section{CHROMOSPHERIC Ly$\alpha$ FLUXES}

     The H~I Ly$\alpha$ line is one of the strongest and most important
emission lines from the upper chromospheres of cool stars.  However, the
very strong ISM absorption that is always present in observations of this
line makes measuring reliable line fluxes difficult (Landsman \& Simon
1993).  In order to measure
an accurate line flux, a high resolution spectrum must be obtained that
resolves the line profile, and then the intrinsic stellar profile must be
reconstructed using the kind of time-consuming analysis technique described
in \S4.  Due to these complications, more attention has been given to other
chromospheric diagnostics, such as the Mg~II h \& k lines.  Since
relatively little has been done with Ly$\alpha$ measurements in the past,
we believe it is worthwhile to see how our ISM-corrected Ly$\alpha$ fluxes
correlate with other, more popular activity diagnostics.

     We use the reconstructed Ly$\alpha$ profiles shown in Figures~6 and 7
to measure integrated Ly$\alpha$ fluxes for our stars, and these fluxes are
listed in Table~3 (in ergs~cm$^{-2}$~s$^{-1}$), along with fluxes measured
from previously analyzed data
(see references in Table~1).  For Capella (\#13) and HR~1099 (\#21), the
fluxes are divided between the two unresolved stars of these binaries.  The
division was made during the absorption analysis for Capella \citep{jll95a},
while for HR~1099 we assume the division between the stars is the
same for Ly$\alpha$ as it is for the Mg~II lines \citep{bew96b}.  Many
of the Ly$\alpha$ profiles in Figures 6 and 7 have self-reversals near line
center.  This is due to the frequent use of the Mg~II h \& k lines to
provide an initial guess for the shape of the Ly$\alpha$ profile.  However,
it should be emphasized that the exact line shape should not be considered
to be terribly precise, especially the parts of the line profile that lie
above the saturated H~I absorption core where the data provide absolutely
no information on the profile.  Uncertainties in the integrated line fluxes
in Table~3 will generally be larger for the lines of sight with higher ISM
column densities since the correction for absorbed flux will generally
be higher in those cases, but a typical average error for the fluxes in
Table~3 is of order 20\%.

     Coronal X-ray and chromospheric Mg~II fluxes are also listed in
Table~3 for comparison with the Ly$\alpha$ fluxes.  The X-ray fluxes are
from the {\em ROSAT} all-sky survey (RASS), which used the Position
Sensitive Proportional Counter (PSPC) instrument.  The exception is
$\chi$~Her (\#43), for which a pointed PSPC observation is used since this
star is undetected in the RASS.  Over half the X-ray fluxes are taken from
the nearby star survey of \citet{js04}.  The bright and giant
star RASS surveys of \citet{mh98a,mh98b} provide many
additional fluxes.  Measurements for three of the Hyades stars (\#55--\#57)
are from \citet{ras95}.  For four stars (\#18, \#19,
\#25, and \#52), we compute a flux ourselves from the RASS count rate
listed in the HEASARC database and an assumed conversion factor of
$6\times 10^{-12}$ ergs~cm$^{-2}$~ct$^{-1}$.  The white dwarf companions of
V471~Tau (\#58) and HD~32008 (\#61) presumably dominate their X-ray fluxes,
so no fluxes are listed in Table~3 for these stars.

     Many binaries are not resolved in {\em ROSAT} PSPC observations,
making it impossible to know how to divide the X-ray flux among the
individual stars.  For the $\alpha$~Cen and 61~Cyg binaries, {\em ROSAT}
High Resolution Imager (HRI) observations that resolve the binaries are
used to determine how to divide the RASS X-ray fluxes among the stars in
Table~3 \citep{js04}.  For Capella (\#13), we assume the
relative contributions of the two stars to the coronal Fe~XXI $\lambda$1356
line \citep{jll98b} are applicable to the coronal X-ray flux as well.
For HR~1099 (\#21) and $\xi$~Boo (\#34), the emission is known to be
dominated by the primary stars of these binaries, so we simply assign all
the X-ray flux to those stars.  Finally, we have recently used the
{\em Chandra} X-ray telescope to resolve the 70~Oph (\#33) and 36~Oph
(\#10) binaries.  These data will be discussed in detail in a future paper,
but we can report here that 70~Oph~A and 36~Oph~A contribute 60.4\% and
58.2\% of the fluxes of these binaries, respectively, so these results are
used to modify the RASS fluxes in Table~3.

     The Mg~II fluxes in Table~3 are combined fluxes of the h \& k lines.
Most of the fluxes are measured from HST spectra.  Although many of the
ISM Mg~II lines are discussed and analyzed by RL02, and some of the
references in Table~1 also discuss Mg~II spectra, emission line fluxes have
not usually been reported so we generally have to measure these fluxes
ourselves.  However, we use the measurements of \citet{jll95b} and
\citet{bew96b} for Capella (\#13) and HR~1099 (\#21), respectively.
The chromospheric Mg~II emission line lies in the middle of a broad
photospheric absorption line.  It is difficult to know what the profile of
the absorption line is underneath the chromospheric emission line.  However,
we measure our Mg~II fluxes by simply integrating the fluxes of the
emission line without trying to correct for any continuum that may be
underneath the lines \citep[see, e.g.,][]{bew96b,bew00a}.  This basically
assumes that the photospheric absorption is broad and saturated underneath
the line.  This assumption will be a significant source of uncertainty
($>10$\%) only for the F stars in our sample, which have the strongest
photospheric UV continua.

     For those stars without high or moderate resolution HST spectra of
Mg~II, we look for high resolution IUE spectra that can be used to measure
Mg~II fluxes.  Fluxes measured from IUE data are flagged in Table~3.  In
all our Mg~II measurements we try to remove ISM absorption before making
the flux measurements.  This is difficult in high resolution IUE or
moderate resolution HST spectra, since the narrow ISM lines are not well
resolved.  It is impossible in low resolution IUE spectra, which do not
even have sufficient resolution to separate the two Mg~II lines, so low
resolution IUE spectra are not used here.

     In order to convert measured fluxes to surface fluxes, we need to
know the radii of our stars.  Stellar radii are therefore
listed in Table~3.  We search the literature for radii that are measured
via interferometry or by other direct means, and references in Table~3
indicate which radii come from such sources.  However, we estimate most
of our radii following the prescription of \citet{bew94}, using
the Barnes-Evans relation for most of the stars \citep{tgb78},
but the radius-luminosity relation of \citet{asg74}
for the M dwarfs.

     Finally, we wish to correlate our Ly$\alpha$ fluxes with rotation
rates, so Table~3 also lists
rotation periods found in the literature for our sample of stars.
References in the table indicate the sources of the periods.  Most are
genuine photometric periods, but some are estimated from Ca~II
measurements and a known Ca~II-rotation relation.  The periods derived
from Ca~II are flagged in Table~3.

     In Figures~15a-b, we plot Ly$\alpha$ surface fluxes versus X-ray
and Mg~II surface fluxes, and in Figure~15c the Ly$\alpha$ fluxes are
plotted against rotation period.  In addition to plotting the stellar data
listed in Table~3, we also add a solar data point to these figures.  We
assume a solar Ly$\alpha$ flux of
$F_{Ly\alpha}=3.5\times 10^{5}$ ergs cm$^{-2}$ s$^{-1}$,
the midpoint of a range quoted by \citet{tnw00}.  A figure in
\citet{tnw04} implies that quiescent solar Mg~II k fluxes are about
1.6 times larger than Ly$\alpha$.  Given that the h line has about 75\%
the flux of the k line, we assume that the total Mg~II h \& k flux is 2.8
times that quoted above for Ly$\alpha$, $F_{MgII}=9.8\times 10^{5}$
ergs cm$^{-2}$ s$^{-1}$.  Finally, the solar rotation period at the
equator is about 26 days, and for the X-ray flux we assume a typical value
of $F_{X}=3.25\times 10^{4}$ ergs cm$^{-2}$ s$^{-1}$ for the Sun
\citep{am87}.

     The chromospheric Ly$\alpha$ fluxes are highly correlated with the
chromospheric Mg~II and coronal X-ray fluxes in Figures~15a-b, as expected.
They are anti-correlated with rotation period in Figure~15c, consistent
with many previous studies showing how rapid rotation induces higher
degrees of chromospheric and coronal activity in cool stars
\citep[e.g.,][]{mm95}.  The Sun is nicely consistent with the
other G star data points in Figures~15a-c, which is encouraging.
Figures 15a-c show separate least squares fits to the F~V+G~V, K~V, and
giant (III) stars, in addition to fits to all the stars combined.  We
combine the F~V stars with the G~V stars since the few F~V stars are
all late F stars and these data points are basically consistent with
the more numerous G stars.  There are not enough subgiants (IV) and M~V
stars to justify making separate fits to those data.  In order
to estimate uncertainties in the fit parameters, we use a Monte Carlo
method.  For a large number of trials, we randomly
vary the data points within assumed error bars, conduct new fits for each
of these trials, and see how much the fit parameters vary.

     The assumed error bars are estimated as follows.  The biggest source
of uncertainty in the fluxes is generally target variability rather than
measurement uncertainty.  \citet{tra97} estimates that {\em ROSAT} PSPC
fluxes should cover a range of about a factor of 4 during solar-like
activity cycles, so we assume 0.3~dex uncertainties for the $F_{X}$ values.
The solar Ly$\alpha$ flux range quoted by \citet{tnw00} represents
a $\pm 20$\% variation from the median value due to solar cycle and
rotational variability.  We assume this result roughly applies to the
Mg~II fluxes as well and therefore estimate 0.08~dex uncertainties for
$F_{MgII}$.  Combining this uncertainty with the previously quoted
$\approx 20$\% measurement uncertainty for the Ly$\alpha$ fluxes leads
us to assume 30\% (0.11~dex) uncertainties for $F_{Ly\alpha}$.  Probably
the major source of uncertainty for the rotation periods is the
presence of differential rotation, which leads to $P_{rot}$ being a
somewhat ill-defined parameter.  We conservatively assume 20\% (0.08~dex)
errors for this quantity.  Table~4 lists the resulting fit parameters
and 1$\sigma$ uncertainties for all the fits in Figures~15a-c.

     Although Ly$\alpha$ and Mg~II are both broadly called chromospheric
lines, the relations between $F_{Ly\alpha}$ and $F_{MgII}$ in Figure~15b
are not quite linear.  For example, the power law index for the F~V+G~V
stars quoted in Table~4 is $0.82\pm 0.04$.  This is well below 1 and
suggests that the Ly$\alpha$ flux increases more rapidly with
activity than Mg~II.  This effect is also evident when comparing our
$F_{Ly\alpha}-F_{X}$ relations with the $F_{MgII}-F_{X}$ relations
from \citet{tra95}.  For solar-like G dwarfs, we find
$F_{X}\propto F_{Ly\alpha}^{2.20\pm 0.13}$ while \citet{tra95}
quotes $F_{X}\propto F_{MgII}^{2.85\pm 0.17}$.  Since higher temperature
lines are known to exhibit stronger increases when plotted as a function of
any activity diagnostic, these results suggest that the Ly$\alpha$ line is
formed at somewhat hotter temperatures than Mg~II, consistent with the
predictions of atmospheric models \citep{jev81}.
It is interesting that the $F_{Ly\alpha}-F_{MgII}$ relation of the giant
stars appears to be noticeably more linear than for the dwarfs, and
surprisingly tight considering the diverse makeup of this group of stars.

     \citet{efg03} quote a power law index of $-1.49$
in their plot of C~II fluxes versus $P_{rot}$ for a selection of
solar-like G stars.  This is somewhat steeper than our analogous
$F_{Ly\alpha}-P_{rot}$ relation, which has a power law of $-1.09\pm 0.08$
(see Table~4).  This suggests that the Ly$\alpha$ lines are formed at
temperatures cooler than C~II, and therefore somewhere in between the Mg~II
and C~II line formation temperatures.

     The K dwarf relations in Figures~15a and 15c are reasonably tight,
but this is not the case in 15b.  The K star data points in Figure~15b
show a lot of scatter, with the two largest discrepancies
being the two latest type stars, 61~Cyg~A and $\epsilon$~Ind (both
K5~V).  Given that the two M dwarfs in the figure are also discrepant in
the same sense, perhaps the relation between Mg~II and Ly$\alpha$
undergoes a drastic change for stars with spectral types later than
early K.  It is also interesting to note that the K dwarf power law
relation in Figure~15a is noticeably steeper than that of the G star
relation, while in Figure~15c it is noticeably shallower.  The implication
is that for a given increase in rotation, the chromospheric and coronal
emission from K stars increases less than that of G stars, but for a given
increase in chromospheric emission the X-ray fluxes of K stars increase
{\em more} than the G stars.

     The last panel of Figure~15 compares our HST Ly$\alpha$ fluxes
with previous measurements from IUE data.  Analyzing the Ly$\alpha$ line
in IUE spectra is much more difficult than in HST spectra for many
reasons.  The S/N is lower, the spectral resolution is poorer, and not only
must one correct for ISM absorption but also for geocoronal
contamination, which can be severe due to the much larger apertures of IUE
compared with HST.  \citet{wl93} developed techniques to
deal with these difficulties and the IUE measurements plotted in
Figure~15d are from their work.  Despite the complications with the IUE
analysis, the agreement between the IUE and HST fluxes is reasonably good.
Some of the scatter could be due to real stellar variability.  For the
lower fluxes, the IUE measurements tend to be systematically higher than
suggested by HST, which may be due to inaccuracies in the IUE's flux
calibration at Ly$\alpha$ or due to the difficulties in subtracting the
geocoronal signal.  The only data point in Figure~15d that is
truly discrepant is EV~Lac, which has an IUE flux about an order of
magnitude larger than the HST flux.  As this M dwarf is a well known
flare star, perhaps there was a strong flare during the IUE observation.

\section{SUMMARY}

     We have analyzed 33 HST/STIS E140M Ly$\alpha$ spectra of cool stars
located within 100~pc, in order to measure chromospheric Ly$\alpha$ fluxes,
determine line-of-sight interstellar H~I and D~I properties, and
search for detections of heliospheric and astrospheric absorption.  Our
findings are summarized as follows.
\begin{description}
\item[1.] We find that the geocoronal Ly$\alpha$ emission lines in our
  spectra tend to be blueshifted with respect to their expected locations
  by $-1.20\pm 0.90$ km~s$^{-1}$, indicating a systematic error in the
  wavelength calibration of STIS E140M spectra, at least in the Ly$\alpha$
  wavelength region.
\item[2.] Our analysis of the D~I and H~I Ly$\alpha$ absorption lines
  has led to 4 new detections of heliospheric Ly$\alpha$ absorption
  (70~Oph, $\xi$~Boo, 61~Vir, and HD~165185), bringing the total number
  of detections to 8.  We find 7 new astrospheric detections
  (EV~Lac, 70~Oph, $\xi$~Boo, 61~Vir, $\delta$~Eri, HD~128987, and DK~UMa),
  although we consider the 61~Vir and DK~UMa detections to be marginal.
  This brings the total number of detected astrospheres to 13.
\item[3.] We have measured interstellar H~I column densities for all 33
  lines of sight, which are listed in Tables~1 and 2.  Of particular
  interest are the surprisingly high column densities towards AD~Leo
  [$\log N({\rm H~I})=18.47$] and HD~203244 [$\log N({\rm H~I})=18.82$].
  For the short 4.69~pc line of sight to AD~Leo, the suggested average
  density of $n({\rm H~I})=0.204$ cm$^{-3}$ represents a new high for a
  Local Bubble line of sight.  The rarity of $\log N({\rm H~I})> 18.8$
  lines of sight within the Local Bubble implies that detections of column
  densities this high are probably due to compact clouds that likely have
  densities well above the $n({\rm H~I})\approx 0.1$ cm$^{-3}$ values
  believed to apply to the very local ISM within 10~pc.
\item[4.] When plotting ISM H~I column densities
  versus $\theta$, the angle between the line of sight and the upwind
  direction of the ISM flow, the heliospheric detections are nicely
  separated from the nondetections.  Only lines of sight with low ISM H~I
  column densities have detected heliospheric absorption, and it is much
  easier to detect the heliospheric absorption in upwind directions (i.e.,
  $\theta<90^{\circ}$). In the most upwind directions, an ISM column of
  $\log N({\rm H~I})< 18.2$ is required to detect heliospheric absorption,
  while in downwind directions one must have $\log N({\rm H~I})< 17.8$.
  This behavior is consistent with expectations from heliospheric models
  and represents further strong evidence in support of the heliospheric
  interpretation of the excess red-side Ly$\alpha$ absorption.
\item[5.] Within 10~pc, 10 of the 17 analyzed HST lines of sight have
  yielded detections of astrospheric absorption, for a detection rate of
  58.8\%.  Since the presence of H~I in the surrounding ISM is one
  necessary ingredient for an astrospheric detection, the 58.8\% detection
  fraction represents a lower limit for the filling factor of neutral ISM
  material within 10~pc, meaning that neutral hydrogen must be present in
  most of the ISM within 10~pc.  However, beyond 10~pc only 3 of 31 HST
  lines of sight have astrospheric detections, for a much lower detection
  rate of 9.7\%.  Some of this difference is due to generally higher ISM
  H~I column densities for longer lines of sight, which results in broader
  ISM absorption that can hide the astrospheric signal.  However, beyond
  10~pc most of the stars are probably surrounded by the hot, ionized ISM
  material that is believed to fill most of the Local Bubble, and this
  will account for many nondetections of astrospheric absorption
  beyond 10~pc.
\item[6.] Our Ly$\alpha$ absorption analyses requires the reconstruction
  of intrinsic stellar Ly$\alpha$ emission lines.  We use these lines to
  measure chromospheric H~I Ly$\alpha$ fluxes for our sample of stars,
  which are corrected for ISM absorption.  We correlate these fluxes with
  coronal X-ray and chromospheric Mg~II fluxes, and we also study the
  dependence of Ly$\alpha$ emission on rotation period.  We present in
  Table~4 the derived relations for various classes of stars.
\end{description}

\acknowledgments

     Support for this work was provided by grant AR-09957 from the Space
Telescope Science Institute, which is operated by AURA, Inc., under
NASA contract NAS5-26555.

\clearpage

\begin{deluxetable}{cccccccccccc}
\tabletypesize{\scriptsize}
\tablecaption{Line of Sight Information for HST Ly$\alpha$ Targets}
\tablecolumns{12}
\tablewidth{0pt}
\tablehead{
  \colhead{ID \#} & \colhead{HD \#} & \colhead{Other} & \colhead{Spectral} &
    \colhead{Dist.} & \colhead{l} & \colhead{b} &
    \colhead{$\log N({\rm H~I})$} & \colhead{Mg~II?} &
    \colhead{Hel.} & \colhead{Ast.} & \colhead{Ref.} \\
  \colhead{} & \colhead{} & \colhead{Name} & \colhead{Type} & \colhead{(pc)} &
    \colhead{(deg)} & \colhead{(deg)} & \colhead{} &\colhead{} &
    \colhead{Det.?} & \colhead{Det.?} &\colhead{}}
\startdata
\multicolumn{12}{c}{PREVIOUS ANALYSES} \\
\hline 
 1&       &Proxima Cen &M5.5 V & 1.30 & 313.9 & -1.9 & 17.61& No&Yes & No&1\\
 2&128620 &$\alpha$ Cen A&G2 V & 1.35 & 315.7 & -0.7 & 17.61&Yes&Yes &Yes&2\\
 3&128621 &$\alpha$ Cen B&K0 V & 1.35 & 315.7 & -0.7 & 17.61&Yes&Yes &Yes&2\\
 4& 48915 & Sirius     &  A1 V & 2.64 & 227.2 & -8.9 & 17.53&Yes&Yes &...&3\\
 5& 22049 &$\epsilon$ Eri&K1 V & 3.22 & 195.8 &-48.1 & 17.88&Yes& No &Yes&4\\
 6&201091 & 61 Cyg A   &  K5 V & 3.48 &  82.3 & -5.8 & 18.13&Yes& No &Yes&5\\
 7& 61421 & Procyon    &F5 IV-V& 3.50 & 213.7 & 13.0 & 18.06&Yes&... &...&6\\
 8&209100 &$\epsilon$ Ind&K5 V & 3.63 & 336.2 &-48.0 & 18.00&No & No &Yes&7\\
 9& 26965 & 40 Eri A   &  K1 V & 5.04 & 200.8 &-38.1 & 17.85&Yes& No & No&5\\
10&155886 & 36 Oph A   &  K1 V & 5.99 & 358.3 &  6.9 & 17.85&Yes&Yes &Yes&8\\
11& 62509 &$\beta$ Gem & K0 III& 10.3 & 192.2 & 23.4 & 18.26&Yes& No & No&4\\
12& 17925 & EP Eri     &  K2 V & 10.4 & 192.1 &-58.3 & 18.05& No&... &...&9\\
13& 34029 & Capella &G8 III+G1 III&12.9&162.6 &  4.6 & 18.24&Yes& No & No&6\\
14&   432 &$\beta$ Cas & F2 IV & 16.7 & 117.5 & -3.3 & 18.13&Yes& No & No&4\\
15& 82443 & DX Leo     & K0 V  & 17.7 & 201.2 & 46.1 & 17.70& No&... &...&9\\
16& 82558 & LQ Hya     & K2 V  & 18.3 & 244.6 & 28.4 & 19.05& No&... &...&9\\
17& 11443 &$\alpha$ Tri& F6 IV & 19.7 & 138.6 &-31.4 & 18.33&Yes& No & No&4\\
18&220140 & V368 Cep   & K2 V  & 19.7 & 118.5 & 16.9 & 17.95& No&... &...&9\\
19&  1405 & PW And     & K2 V  & 21.9 & 114.6 &-31.4 & 18.35& No&... &...&9\\
20&222107&$\lambda$ And&G8 IV-III+M V&25.8&109.9&-14.5&18.45& No& No&Yes?&7\\
21& 22468 & HR 1099&K1 IV+G5 IV& 29.0 & 184.9 &-41.6 & 18.13&Yes& No &No&10\\
22&  4128 &$\beta$ Cet & K0 III& 29.4 & 111.3 &-80.7 & 18.36&Yes&...&...&10\\
23&       & HZ 43      &   DA  & 32.0 &  54.1 & 84.2 &17.93&Yes&Yes?&...&11\\
24& 62044 &$\sigma$ Gem& K1 III& 37.5 & 191.2 & 23.3 & 18.20&Yes& No & No&4\\
25&197890 & Speedy Mic & K2-3 V& 44.4 &   6.3 &-38.3 & 18.30& No&... &...&9\\
26&       & G191-B2B   &   DA  & 68.8 & 156.0 &  7.1 & 18.18&Yes& No&...&12\\
27&       & Feige 24   &   DA  & 74.4 & 166.0 &-50.3 & 18.47& No& No&...&13\\
28&       & GD 246     &   DA  & 79.0 &  87.3 &-45.1 & 19.11& No& No&...&14\\
29&111812 & 31 Com     & G0 III& 94.2 & 115.0 & 89.6 & 17.88&Yes& No & No&4\\
\hline 
\multicolumn{12}{c}{NEW ANALYSES\tablenotemark{a}} \\
\hline 
30& 10700 & $\tau$ Cet &  G8 V & 3.65 & 173.1 &-73.4 & 18.01&No & No &No&15\\
31&       & AD Leo     &M3.5 V & 4.69 & 216.5 & 54.6 & 18.47&No & No &No&15\\
32&       & EV Lac     &M3.5 V & 5.05 & 100.6 &-13.1 & 17.97&No & No&Yes&15\\
33&165341 & 70 Oph A   &  K0 V & 5.09 &  29.9 & 11.4 & 18.06&Yes&Yes&Yes&15\\
34&131156A& $\xi$ Boo A&  G8 V & 6.70 &  23.1 & 61.4 & 17.92&Yes&Yes&Yes&15\\
35&115617 & 61 Vir     &  G5 V & 8.53 & 311.9 & 44.1 & 17.91&No&Yes&Yes?&15\\
36& 39587 &$\chi^1$ Ori&  G0 V & 8.66 & 188.5 & -2.7 & 17.93&Yes& No &No&15\\
37& 23249 &$\delta$ Eri& K0 IV & 9.04 & 198.1 &-46.0 & 17.88&No & No&Yes&15\\
38& 20630 &$\kappa$ Cet&  G5 V & 9.16 & 178.2 &-43.1 & 17.89&Yes& No &No&15\\
39&197481 & AU Mic     &  M0 V & 9.94 &  12.7 &-36.8 & 18.36&Yes& No &No&15\\
40& 33262 &$\zeta$ Dor &  F7 V & 11.7 & 266.0 &-36.7 & 18.29&Yes& No &No&15\\
41& 37394 & HR 1925    &  K1 V & 12.2 & 158.4 & 12.0 & 18.26&No & No &No&15\\
42&   166 & HR 8       &  K0 V & 13.7 & 111.3 &-32.8 & 18.29&No & No &No&15\\
43&142373 &$\chi$ Her  &  F8 V & 15.9 &  67.7 & 50.3 & 18.25&No & No &No&15\\
44& 43162 & HR 2225    &  G5 V & 16.7 & 230.9 &-18.5 & 17.87&No & No &No&15\\
45&165185 & HR 6748    &  G5 V & 17.4 & 356.0 & -7.3 & 18.13&No &Yes &No&15\\
46&203244 &SAO 254993  &  G5 V & 20.5 & 324.9 &-38.9 & 18.82&No & No &No&15\\
47& 97334 & HR 4345    &  G0 V & 21.7 & 184.3 & 67.3 & 17.82&No & No &No&15\\
48& 59967 & HR 2882    &  G4 V & 21.8 & 250.5 & -9.0 & 18.54&No & No &No&15\\
49&116956 &SAO 28753   &G9 IV-V& 21.9 & 113.7 & 59.5 & 18.19&No & No &No&15\\
50&106516 & HR 4657    &  F5 V & 22.6 & 288.5 & 51.5 & 18.57&No & No &No&15\\
51& 73350 &SAO 136111  &  G0 V & 23.6 & 232.1 & 20.0 & 18.16&No & No &No&15\\
52&128987 &SAO 158720  &  G6 V & 23.6 & 337.5 & 39.2 & 18.11&No & No&Yes&15\\
53& 82210 & DK UMa     &G4 III-IV&32.4& 142.6 & 38.9 & 18.05&Yes&No&Yes?&15\\
54& 93497 &$\mu$ Vel   &G5 III & 35.3 & 283.0 &  8.6 & 18.47&No & No &No&15\\
55& 28568 &SAO 93981   &  F2 V & 41.2 & 180.5 &-21.4 & 18.12&Yes& No &No&15\\
56& 28205 &V993 Tau    &  G0 V & 45.8 & 180.4 &-22.4 & 18.11&Yes& No &No&15\\
57& 28033 &SAO 76609   &  F8 V & 46.4 & 175.4 &-18.9 & 18.15&Yes& No &No&15\\
58&       &V471 Tau    &  K0 V & 46.8 & 172.5 &-27.9 & 18.21&No & No &No&15\\
59&209458 &V376 Peg    &  G0 V & 47.1 &  76.8 &-28.5 & 18.37&No & No &No&15\\
60&220657 &$\upsilon$ Peg&F8 III&53.1 &  98.6 &-35.4 & 18.24&Yes& No &No&15\\
61& 32008 & HR 1608  &G7 IV-III& 54.7 & 209.6 &-29.4 & 18.08&No & No &No&15\\
62&203387 &$\iota$ Cap &G8 III & 66.1 &  33.6 &-40.8 & 18.70&Yes& No &No&15\\
\enddata
\tablenotetext{a}{All new analyses are from HST/STIS E140M spectra,
  with a resolution of $R\approx 45,000$.}
\tablerefs{(1) Wood et al.\ 2001. (2) Linsky \& Wood 1996. (3) Bertin et al.\
  1995. (4) Dring et al.\ 1997. (5) Wood \& Linsky 1998. (6) Linsky et al.\
  1995a. (7) Wood et al.\ 1996a. (8) Wood et al.\ 2000b. (9) Wood et al.\
  2000a. (10) Piskunov et al.\ 1997. (11) Kruk et al.\ 2002. (12) Lemoine
  et al.\ 2002. (13) Vennes et al.\ 2000. (14) Oliveira et al.\ 2003. (15)
  This paper.}
\end{deluxetable}

\clearpage

\begin{deluxetable}{cccccccccc}
\tabletypesize{\scriptsize}
\tablecaption{Ly$\alpha$ Fit Parameters}
\tablecolumns{10}
\tablewidth{0pt}
\tablehead{
  \colhead{ID \#} & \colhead{$V_{LIC}$} & \colhead{$V_{G}$} &
    \colhead{Comp.\tablenotemark{a}} &
    \colhead{$v({\rm Mg~II})$\tablenotemark{b}} &
    \colhead{$v({\rm H~I})$\tablenotemark{c}} &
    \colhead{$b({\rm H~I})$\tablenotemark{c}} &
    \colhead{$\log N({\rm H~I})$\tablenotemark{c}} &
    \colhead{$\chi^{2}_{\nu}$} & \colhead{Notes\tablenotemark{d}} \\
  \colhead{} & \colhead{(km~s$^{-1}$)} & \colhead{(km~s$^{-1}$)} &\colhead{}&
    \colhead{(km~s$^{-1}$)} & \colhead{(km~s$^{-1}$)} &
    \colhead{(km~s$^{-1}$)} & \colhead{} & \colhead{} &\colhead{}}
\startdata
30& 13.8& 17.6& 1& ...  & $12.34\pm 0.06$&$10.32\pm 0.06$&$18.006\pm 0.002$&
  1.106&  \\
31&  6.4&  5.2& 1& ...  & $13.13\pm 0.05$&$14.05\pm 0.06$&$18.475\pm 0.001$&
  2.571& *\\
32&  3.5&  5.2& 1& ...  &  $7.3\pm 0.3$  & $11.1\pm 0.2$ &$17.972\pm 0.007$&
  0.820& *\\
  &     &     &HS/AS&...& $-6.7\pm 2.2$  & $29.1\pm 1.2$ & $14.46\pm 0.07$ &
       &  \\
33&-23.5&-26.4& 1&-26.64& $-26.9\pm 0.2$ &  $8.7\pm 0.4$ &$17.934\pm 0.003$&
  0.851& *\\
  &     &     & 2&-32.91&   $-33.22$     &    $8.72$     &     $17.293$    &
       &  \\
  &     &     & 3&-43.36&   $-43.68$     &    $8.72$     &     $16.972$    &
       &  \\
  &     &     &HS/AS&...&$-32.62\pm 0.09$& $26.4\pm 0.5$ & $15.39\pm 0.06$ &
       &  \\
34&-17.7&-21.6& 1&-17.15& $-19.3\pm 0.3$ &  $9.7\pm 0.6$ &$17.915\pm 0.007$&
  1.191& *\\
  &     &     &HS/AS&...& $-16.9\pm 0.3$ & $19.2\pm 0.7$ &  $15.7\pm 0.1$  &
       &  \\
35&-15.4&-19.2& 1& ...  & $-16.5\pm 0.7$ &  $7.4\pm 1.5$ & $17.91\pm 0.04$ &
  0.992& *\\
  &     &     &HS/AS&...& $-12.4\pm 0.9$ & $18.1\pm 1.9$ &  $16.1\pm 0.5$  &
       &  \\
36& 24.9& 27.9& 1& 23.35&  $23.3\pm 0.1$ &$11.95\pm 0.06$&$17.934\pm 0.003$&
  1.251&  \\
37& 22.0& 26.0& 1& ...  &  $19.6\pm 0.2$ & $11.8\pm 0.1$ &$17.880\pm 0.007$&
  1.061& *\\
  &     &     &HS/AS&...&   $2.6\pm 0.9$ & $31.3\pm 0.7$ & $14.56\pm 0.03$ &
       &  \\
38& 22.8& 27.0& 1& 20.78&  $20.0\pm 0.4$ & $11.1\pm 0.2$ &$17.592\pm 0.006$&
  0.967& *\\
  &     &     & 2& 13.33&    $12.51$     &   $11.07$     &     $17.550$    &
       &  \\
  &     &     & 3&  7.25&     $6.43$     & $13.3\pm 0.2$ &     $16.545$    &
       &  \\
39&-15.3&-15.7& 1&-21.22&$-20.83\pm 0.09$&$13.74\pm 0.08$&$18.356\pm 0.002$&
  0.879& *\\
40&  7.8&  9.4& 2&  8.45&  $8.6\pm 0.1$  &  $9.8\pm 0.5$ &$18.090\pm 0.003$&
  1.273& *\\
  &     &     & 1& 14.12&    $14.27$     & $14.8\pm 0.1$ &     $17.870$    &
       &  \\
41& 19.8& 22.1& 1& ...  & $17.5\pm 0.2$  & $11.9\pm 0.3$ &$18.256\pm 0.005$&
  0.791&  \\
42&  9.4& 12.3& 1& ...  &  $6.5\pm 0.2$  & $12.7\pm 0.2$ &$18.290\pm 0.006$&
  0.577&  \\
43&-13.1&-15.9& 1& ...  &$-12.9\pm 0.1$  & $12.4\pm 0.2$ &$18.249\pm 0.003$&
  0.920&  \\
44& 18.9& 21.3& 1& ...  & $14.0\pm 0.2$  & $12.8\pm 0.1$ & $17.87\pm 0.01$ &
  1.106&  \\
45&-23.2&-25.7& 1& ...  & $-29.2\pm 0.7$ & $13.4\pm 0.9$ & $18.13\pm 0.02$ &
  0.938& *\\
  &     &     &HS/AS&...& $-20.8\pm 3.5$ & $17.8\pm 1.8$ &  $16.2\pm 0.5$  &
       &  \\
46& -9.9&-10.0& 1& ...  &$-13.1\pm 0.7$  & $15.9\pm 1.2$ & $18.82\pm 0.01$ &
  1.125& *\\
47&  2.8&  1.1& 1& ...  &  $4.3\pm 0.2$  & $11.7\pm 0.1$ & $17.82\pm 0.01$ &
  1.022&  \\
48& 11.7& 12.7& 1& ...  &  $8.4\pm 0.4$  & $14.0\pm 0.6$ &$18.536\pm 0.008$&
  1.095&  \\
49& -2.5& -4.3& 1& ...  &  $3.2\pm 0.3$  & $13.8\pm 0.2$ &$18.188\pm 0.009$&
  1.499&  \\
50& -9.0&-12.2& 1& ...  & $-3.4\pm 1.1$  & $18.9\pm 0.8$ & $18.57\pm 0.02$ &
  0.991& *\\
51& 13.6& 13.9& 1& ...  & $12.0\pm 0.4$  & $12.8\pm 0.3$ &$18.163\pm 0.009$&
  1.224&  \\
52&-21.4&-25.5& 1& ...  &$-22.0\pm 0.7$  & $13.1\pm 0.4$ & $18.11\pm 0.02$ &
  1.246& *\\
  &     &     &HS/AS&...&   $-42.74$     & $34.1\pm 2.4$ & $14.39\pm 0.06$ &
       &  \\
53&  9.3&  9.5& 1&  9.63& $8.30\pm 0.08$ & $10.1\pm 0.3$ &$18.051\pm 0.002$&
  1.059& *\\
  &     &     &HS/AS&...&  $5.4\pm 0.9$  & $19.6\pm 0.8$ &  $15.2\pm 0.2$  &
       &  \\
54& -4.0& -5.6& 1& ...  &$-4.38\pm 0.05$ &$12.90\pm 0.09$&$18.469\pm 0.001$&
  1.607&  \\
55& 25.5& 29.3& 1& 23.9 & $26.6\pm 0.6$  & $12.4\pm 0.5$ & $18.03\pm 0.02$ &
  0.887&  \\
  &     &     & 2& 16.5 &    $19.21$     &   $12.42$     &    $17.371$     &
       &  \\
56& 25.9& 29.3& 1& 23.3 & $24.1\pm 0.6$  & $12.7\pm 0.3$ & $18.00\pm 0.01$ &
  0.920&  \\
  &     &     & 2& 14.8 &    $15.65$     &   $12.74$     &    $17.471$     &
       &  \\
57& 25.3& 29.1& 1& 23.6 & $23.8\pm 0.8$  & $12.8\pm 0.6$ & $18.15\pm 0.02$ &
  0.860&  \\
58& 24.6& 28.6& 1& ...  & $20.9\pm 0.3$  & $12.3\pm 0.4$ &$18.206\pm 0.007$&
  0.948& *\\
59& -3.7& -2.5& 1& ...  & $-6.6\pm 0.9$  & $12.3\pm 1.3$ & $18.37\pm 0.02$ &
  0.949& *\\
60&  5.1&  7.5& 2&  1.13\tablenotemark{e}& $2.21\pm 0.09$ &$11.62\pm 0.05$&
  $18.157\pm 0.001$& 1.076& *\\
  &     &     & 1&  7.00\tablenotemark{e}&     $8.09$     & $14.0\pm 1.4$ &
      $17.157$     &      &  \\
  &     &     & 3& -7.61\tablenotemark{e}&    $-6.52$     & $13.4\pm 0.2$ &
      $17.167$     &      &  \\
61& 23.3& 26.8& 1& ...  &  $6.1\pm 1.0$  & $12.9\pm 0.3$ & $17.69\pm 0.06$ &
  0.765& *\\
  &     &     & 2& ...  & $21.6\pm 0.9$  &   $12.87$     & $17.84\pm 0.05$ &
       &  \\
62&-11.8&-11.5& 3& -2.17\tablenotemark{e}& $-8.5\pm 0.1$  & $15.9\pm 0.2$ &
  $18.563\pm 0.002$& 1.681& *\\
  &     &     & 2&-12.67\tablenotemark{e}&   $-18.94$     &   $15.91$     &
      $17.903$     &      &  \\
  &     &     & 1&-19.35\tablenotemark{e}&   $-25.63$     &   $15.91$     &
      $17.723$     &      &  \\
\enddata
\tablenotetext{a}{Component number, where we assign the same numbers as
  Redfield \& Linsky (2002) when a correspondence exists. ``HS/AS'' indicates
  an absorption component that accounts for heliospheric and/or astrospheric
  absorption.  It has been demonstrated previously that hydrodynamic modeling
  of the HS/AS absorption is required to truly quantify it properly, so the
  empirical fit parameters quoted here for the HS/AS components should be
  regarded with skepticism.}
\tablenotetext{b}{ISM Mg~II absorption velocity from Redfield \& Linsky
  (2002).}
\tablenotetext{c}{Quoted uncertainties are 1$\sigma$ random errors in the
  fit, which do not include estimates of systematic errors (see \S4.2).}
\tablenotetext{d}{When indicated, see \S4.5 for more information on the
  Ly$\alpha$ analysis.}
\tablenotetext{e}{This is an Fe~II velocity instead of Mg~II.}
\end{deluxetable}

\clearpage

\begin{deluxetable}{cccccccc}
\tabletypesize{\scriptsize}
\tablecaption{Stellar Information for HST Ly$\alpha$ Targets}
\tablecolumns{8}
\tablewidth{0pt}
\tablehead{
  \colhead{ID \#} & \multicolumn{2}{c}{Radius} &
    \multicolumn{2}{c}{$P_{rot}$} & \multicolumn{3}{c}{Fluxes (10$^{-12}$)}\\
  \colhead{} & \colhead{($R_{\odot}$)} & \colhead{Ref.} & \colhead{(days)} &
   \colhead{Ref.} & \colhead{Ly$\alpha$} & \colhead{Mg II} & \colhead{X-ray}}
\startdata
 1 & 0.15 &  1 & 41.6 & 11 & 4.21 & ...  & 8.21 \\
 2 & 1.22 &  2 &   29 & 12 & 101  & 398  & 10.7 \\
 3 & 0.86 &  2 &   43 & 13 & 150  & 285  & 12.1 \\
 5 & 0.78 & ...& 11.7 & 14 & 48.8 & 61.7 & 16.8 \\
 6 & 0.68 & ...& 35.4 & 14 & 17.2 & 14.2 & 1.94 \\
 7 & 2.05 &  3 &  ... & ...& 148  & 512  & 22.1 \\
 8 & 0.75 & ...&   22\tablenotemark{a} &15&31.0&15.1\tablenotemark{b}&1.56 \\
 9 & 0.82 & ...&   42 & 16 & 6.93 & 11.6 & 1.09 \\
10 & 0.69 & ...& 20.7 & 14 & 14.2 & 11.0 & 2.93 \\
11 & 8.83 &  4 &  ... & ...& 42.3 & 98.5 & 0.36 \\
12 & 0.93 & ...& 6.76 & 14 & 6.05 & 13.4 & 7.21 \\
13A& 12.2 &  5 & 106  & 17 & 157  & 414  & 77.8 \\
13B&  9.2 &  5 & 8.64 & 17 & 268  & 683  & 63.2 \\
14 & 3.80 &  6 &  ... & ...& 21.7 & ...  & 0.97 \\
15 & 0.80 & ...& 5.38 & 18 & 2.31 & 5.87 & 4.43 \\
16 & 0.73 & ...& 1.60 & 19 & 4.01 & 5.10 & 16.5 \\
17 & 3.04 & ...& 1.74 & 20 & 38.8 & 115  & 6.07 \\
18 & 0.80 & ...& 2.77 & 21 & 2.99 & 5.52 & 17.5 \\
19 & 0.68 & ...& 1.75 & 22 & 2.31 & 2.54 & 9.14 \\
20 &  7.4 &  6 &   54 & 23 & 75.3 & 171\tablenotemark{b}  & 82.9 \\
21A&  3.9 &  7 & 2.84 &  7 & 94.2 & 51.0 & 150  \\
21B&  1.3 &  7 & 2.84 &  7 & 7.57 &  4.1 & ...  \\
22 & 16.8 &  8 &  ... & ...& 85.1 & 158  & 26.6 \\
24 &  9.3 &  6 & 19.6 & 24 & 74.0 & 107  & 74.8 \\
25 & 0.85 & ...& 0.38 & 25 & 1.85 & 1.64 & 36.7 \\
29 &  8.9 & ...& 6.96 & 24 & 11.1 & 35.4 & 11.6 \\
30 & 0.77 &  9 & 34.5\tablenotemark{a} &15&9.99&14.0\tablenotemark{b}&0.31 \\
31 & 0.38 & ...&  2.6 & 26 & 9.95 & 2.79\tablenotemark{b} & 24.0 \\
32 & 0.35 & ...& 4.38 & 26 & 2.75 & ...  & 32.0 \\
33 & 0.85 & ...& 19.7 & 27 & 21.4 & 27.8 & 6.01 \\
34 & 0.78 & ...& 6.31 & 14 & 18.5 & 32.4 & 14.8 \\
35 & 1.00 & ...&  ... & ...& 1.69 & 4.53 & 0.085\\
36 & 0.98 & ...& 5.36 & 14 & 12.7 & 25.0 & 11.7 \\
37 & 2.58 & ...& 55.3\tablenotemark{a} &27&7.82&16.3\tablenotemark{b}&0.11 \\
38 & 0.99 & ...& 9.24 & 14 & 8.44 & 14.9 & 7.21 \\
39 & 0.61 & ...& 4.85 & 26 & 10.3 & 4.20 & 36.1 \\
40 & 0.96 & ...&  ... & ...& 7.99 & 18.5 & 2.86 \\
41 & 0.88 & ...& 10.9 & 28 & 4.57 & 7.02 & 2.24 \\
42 & 0.90 & ...& 6.23 & 28 & 4.77 & 7.17\tablenotemark{b} & 4.15 \\
43 & 1.56 & ...& 16.8\tablenotemark{a} &27&2.04&4.20\tablenotemark{b}&0.029\\
44 & 0.87 & ...&  ... & ...& 3.45 & ...  & 4.04 \\
45 & 0.94 & ...&  5.9\tablenotemark{a} & 15 & 3.73 & 7.71 & 4.08 \\
46 & 0.89 & ...&  ... & ...& 2.47 & 3.33 & 1.14 \\
47 & 1.01 & ...& 8.25 & 28 & 2.09 & 5.29 & 1.95 \\
48 & 0.89 & ...&  ... & ...& 2.76 & 4.45 & 2.07 \\
49 & 0.91 & ...&  7.8 & 28 & 1.62 & 2.82 & 1.21 \\
50 & 0.96 & ...& 6.91 & 14 & 1.28 & 2.88 & 0.25 \\
51 & 1.07 & ...& 6.14 & 28 & 1.34 & 2.69 & 0.79 \\
52 & 0.84 & ...& 9.35 & 28 & 1.44 & 2.52 & 0.60 \\
53 &  4.4 & ...&   10\tablenotemark{a} & 29 & 15.6 & 33.1 & 18.2 \\
54 & 14.4 & ...&  ... & ...& 55.3 & 131\tablenotemark{b}  & 22.2 \\
55 & 1.35 & ...&  ... & ...& 1.41 & 2.98 & 1.32 \\
56 & 1.16 & ...& 4.65 & 30 & 0.59 & 1.49 & 0.44 \\
57 & 1.20 & ...&  ... & ...& 0.27 & 0.55 & 0.21 \\
58 & 0.96 & 10 & 0.52 & 31 & 2.87 & ...  & ...  \\
59 & 1.10 & ...& 15.7\tablenotemark{a} & 32 & 0.15 & ...  & ...  \\
60 &  6.3 & ...&  ... & ...& 11.8 & 33.4 & 6.95 \\
61 & 5.17 & ...&  ... & ...& 5.03 & ...  & ...  \\
62 & 12.3 & ...&   68 & 33 & 12.9 & 27.1 & 8.54 \\
\enddata
\tablenotetext{a}{Estimated from a Ca~II-rotation
  relation rather than from a photometric periodicity.}
\tablenotetext{b}{Measured from high resolution IUE spectra instead of
  HST data.}
\tablerefs{(1) S\'{e}gransan et al.\ 2003. (2) Kervella et al.\ 2003.
  (3) Kervella et al.\ 2004. (4) Mozurkewich et al.\ 2003. (5) Hummel et al.\
  1994. (6) Nordgren et al.\ 1999. (7) Fekel 1983. (8) Bord\'{e} et al.\
  2002. (9) Pijpers et al.\ 2003. (10) O'Brien et al.\ 2001. (11) Benedict
  et al.\ 1993. (12) Hallam et al.\ 1991. (13) Char et al.\ 1993. (14)
  Donahue et al.\ 1996. (15) Saar \& Osten 1997. (16) Frick et al.\ 2004.
  (17) Strassmeier et al.\ 2001. (18) Messina et al.\ 1999. (19) Jetsu 1993.
  (20) Vilhu \& Rucinski 1983. (21) Mantegazza et al.\ 1992. (22) Hooten \&
  Hall 1990. (23) Jetsu 1996. (24) Strassmeier et al.\ 1999. (25) Cutispoto
  et al.\ 1997. (26) Hempelmann et al.\ 1995. (27) Noyes et al.\ 1984. (28)
  Gaidos et al.\ 2000. (29) Young et al.\ 1989. (30) Paulson et al.\ 2004.
  (31) Nelson \& Young 1970. (32) Barnes 2001. (33) Choi et al.\ 1995.}
\end{deluxetable}

\clearpage

\begin{deluxetable}{cccc}
\tabletypesize{\scriptsize}
\tablecaption{Parameters for Fits in Figure 15}
\tablecolumns{4}
\tablewidth{0pt}
\tablehead{
  \colhead{Equation} & \colhead{Spect.\ Type} & \colhead{$\alpha$} &
    \colhead{$\beta$}}
\startdata
$\log F_{X}=\alpha\log F_{Ly\alpha}+\beta$ & G V+F V & $2.20\pm 0.13$ &
  $-7.76\pm 0.82$ \\
                                           &   K V   & $2.90\pm 0.20$ &
 $-12.11\pm 1.28$ \\
                                           &   III   & $2.23\pm 0.18$ &
  $-7.70\pm 1.06$ \\
                                           &   all   & $2.22\pm 0.08$ &
  $-7.81\pm 0.47$ \\
$\log F_{MgII}=\alpha\log F_{Ly\alpha}+\beta$&G V+F V& $0.82\pm 0.04$ &
  $1.43\pm 0.26$ \\
                                           &   K V   & $0.89\pm 0.06$ &
  $0.82\pm 0.40$ \\
                                           &   III   & $0.98\pm 0.06$ &
  $0.50\pm 0.36$ \\
                                           &   all   & $0.71\pm 0.02$ &
  $2.04\pm 0.14$ \\
$\log F_{Ly\alpha}=\alpha\log P_{rot}+\beta$&G V+F V & $-1.09\pm 0.08$ &
  $7.14\pm 0.08$ \\
                                           &   K V   & $-0.57\pm 0.03$ &
  $6.82\pm 0.03$ \\
                                           &   III   & $-0.43\pm 0.06$ &
  $6.68\pm 0.09$ \\
                                           &   all   & $-0.65\pm 0.02$ &
  $6.82\pm 0.02$ \\
\enddata
\end{deluxetable}

\begin{figure}
\plotfiddle{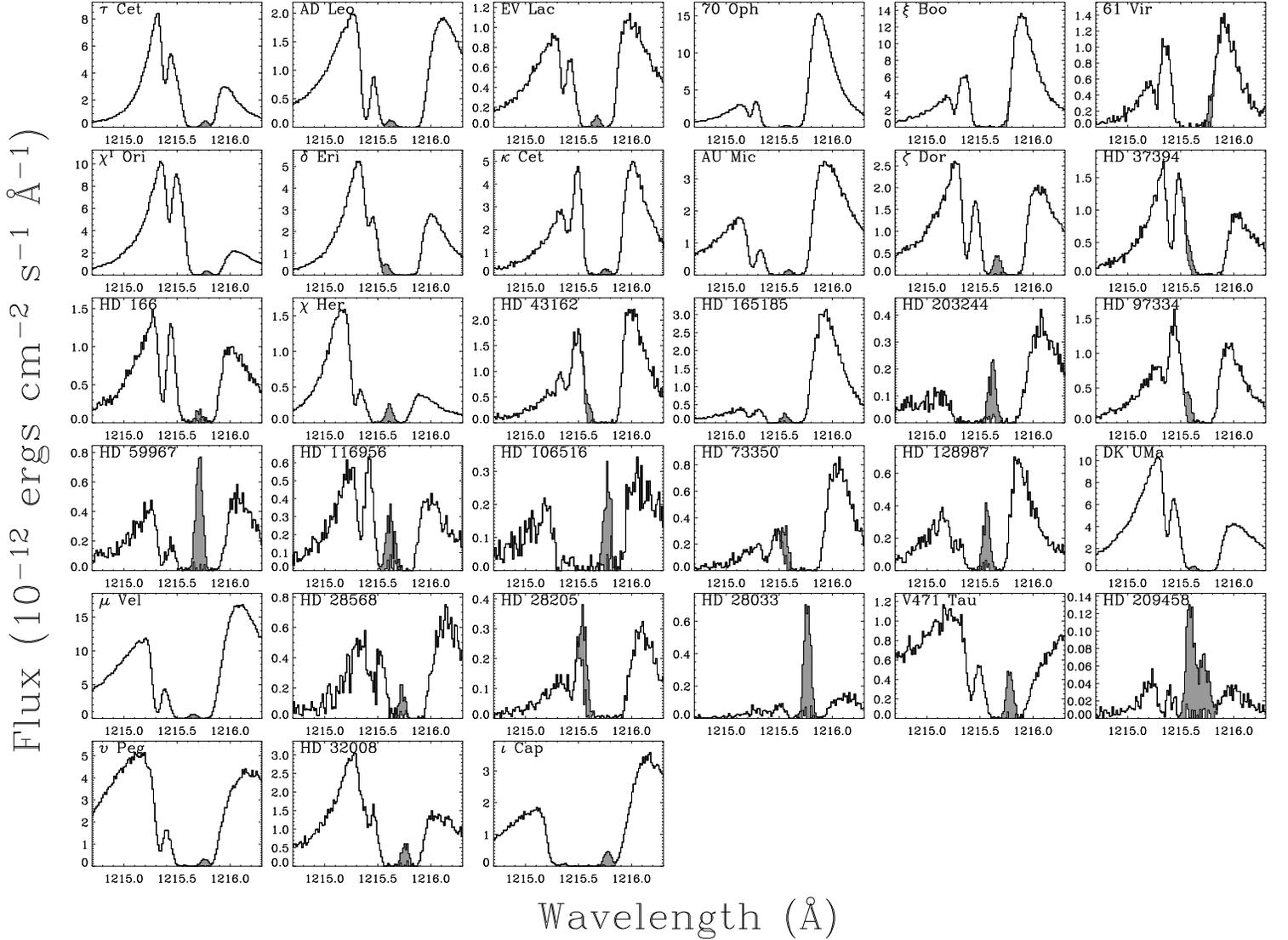}{5.0in}{90}{80}{80}{325}{20}
\caption{Spectra of the H~I Ly$\alpha$ line.  The observations are all
  HST/STIS E140M spectra.  The spectra show very broad, saturated H~I
  absorption from the ISM near line center, and most show weaker and
  narrower D~I absorption $-0.33$~\AA\ from the center of the H~I line.
  The shaded regions are geocoronal Ly$\alpha$ emission, which are
  subtracted from the spectra before the H~I and D~I absorption lines are
  analyzed.}
\end{figure}

\begin{figure}
\plotfiddle{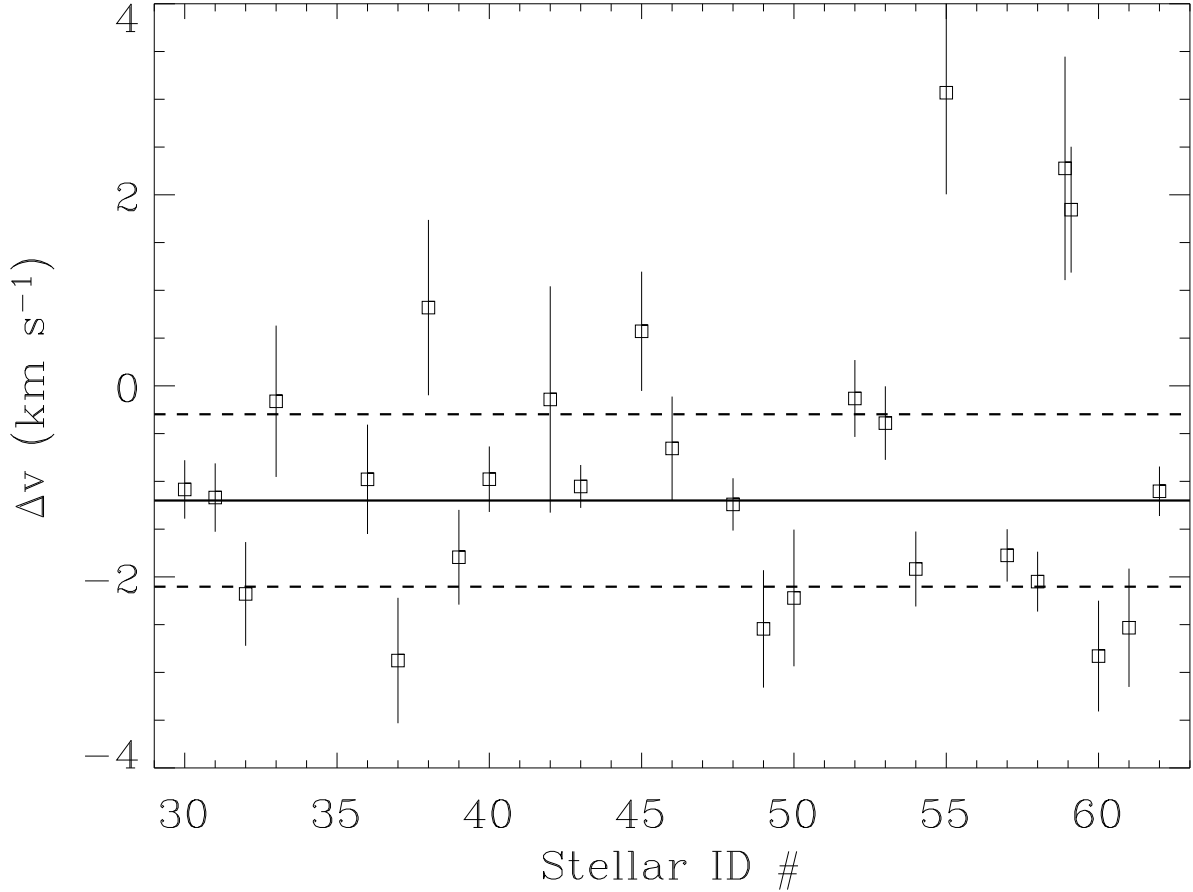}{3.5in}{90}{75}{75}{290}{0}
\caption{We subtract from the measured centroid of the geocoronal
  Ly$\alpha$ emission line its expected velocity (i.e.,
  $\Delta v \equiv v_{\rm obs}-v_{\rm geo}$) and plot the resulting
  velocity discrepancies for all the newly analyzed spectra shown in Fig.~1,
  where the stellar ID numbers refer to those used in Table~1.  We cannot
  measure a centroid in cases where the geocoronal emission is blended with
  the stellar emission, so no point is plotted for some stars.  For star
  \#59 there are two points plotted since there are observations at two
  different times.  The horizontal solid line and the dashed lines show the
  weighted mean and standard deviation of the $\Delta v$ values,
  $\Delta v=-1.20\pm 0.90$ km~s$^{-1}$.}
\end{figure}

\begin{figure}
\plotfiddle{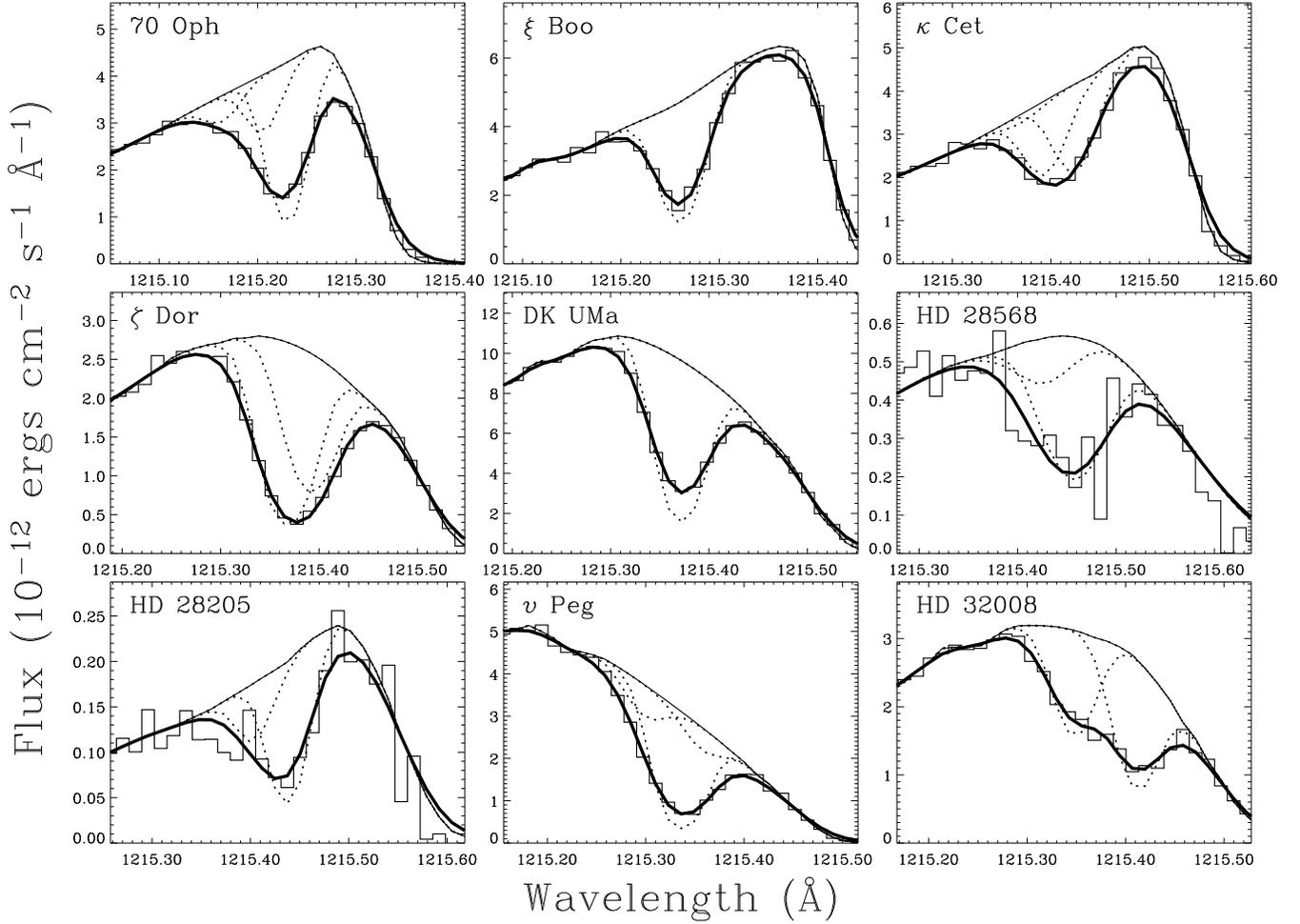}{3.5in}{90}{75}{75}{290}{0}
\caption{Fits to the D~I Ly$\alpha$ absorption line for selected stars.
  Dotted lines show the individual components, and the thick solid
  line shows the total absorption after convolution with the LSF, which
  fits the data.  For the multi-component fits, the fits are constrained
  using information on the ISM velocity structure provided by previous
  analyses of Mg~II and/or Fe~II absorption lines.}
\end{figure}

\begin{figure}
\plotfiddle{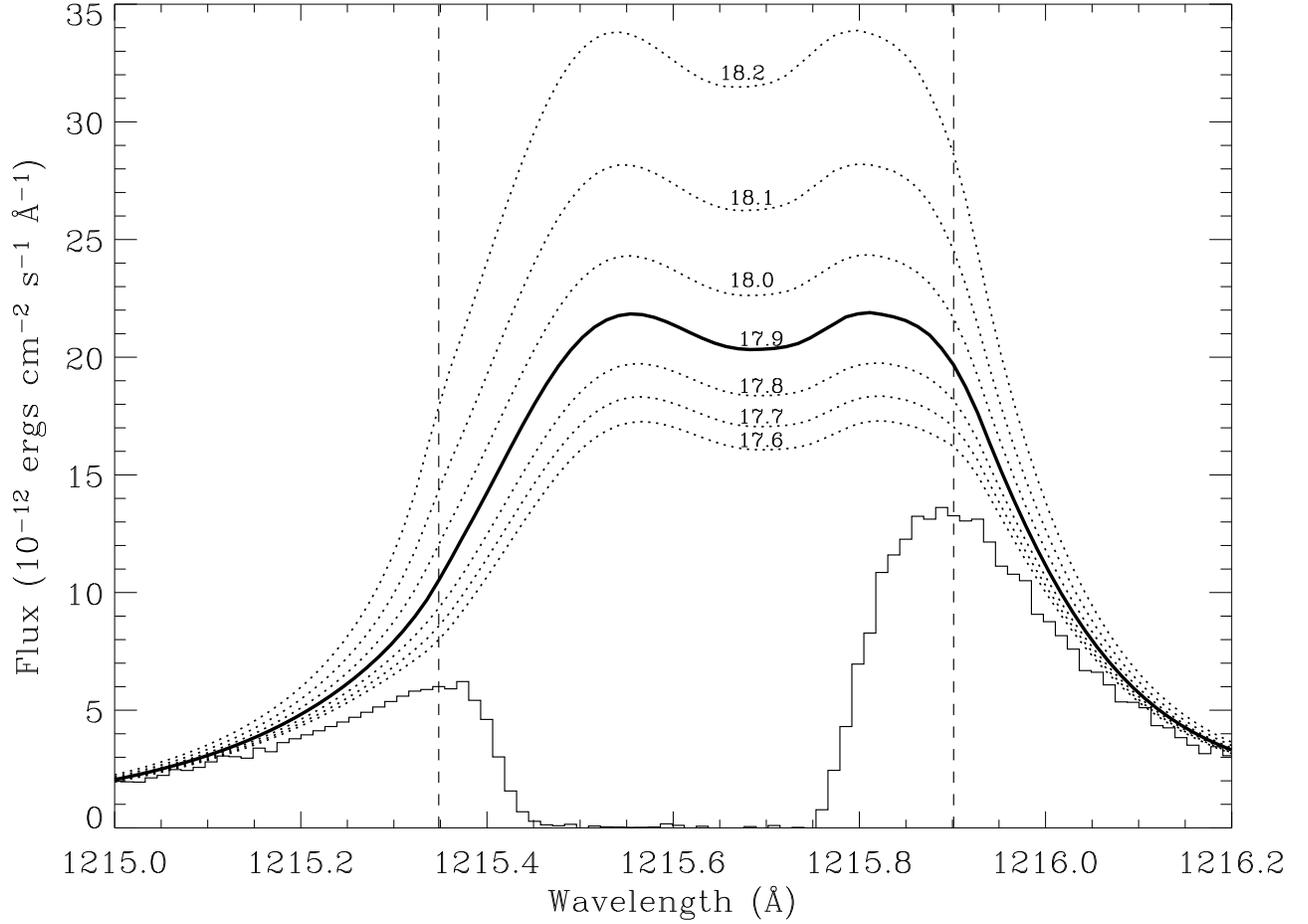}{3.5in}{90}{75}{75}{290}{0}
\caption{Reconstructions of the stellar Ly$\alpha$ emission line
  profile for $\xi$~Boo~A, assuming ISM H~I column densities of
  $\log N({\rm H~I})=17.6-18.2$.  Outside of the dashed lines the profile
  is estimated by extrapolating upwards from the data based on the
  assumed $\log N({\rm H~I})$, while inside of the dashed lines the profile
  shape is estimated from the observed profile of the Mg~II h \& k lines.
  Note that the D~I absorption line at 1215.25~\AA\ has been removed.
  Measurements of the D~I absorption suggest that the H~I column density
  should be $\log N({\rm H~I})=17.9$, assuming a canonical value of
  ${\rm D/H}=1.5\times 10^{-5}$, so the profile constructed assuming
  this column density is emphasized in the figure.}
\end{figure}

\begin{figure}
\plotfiddle{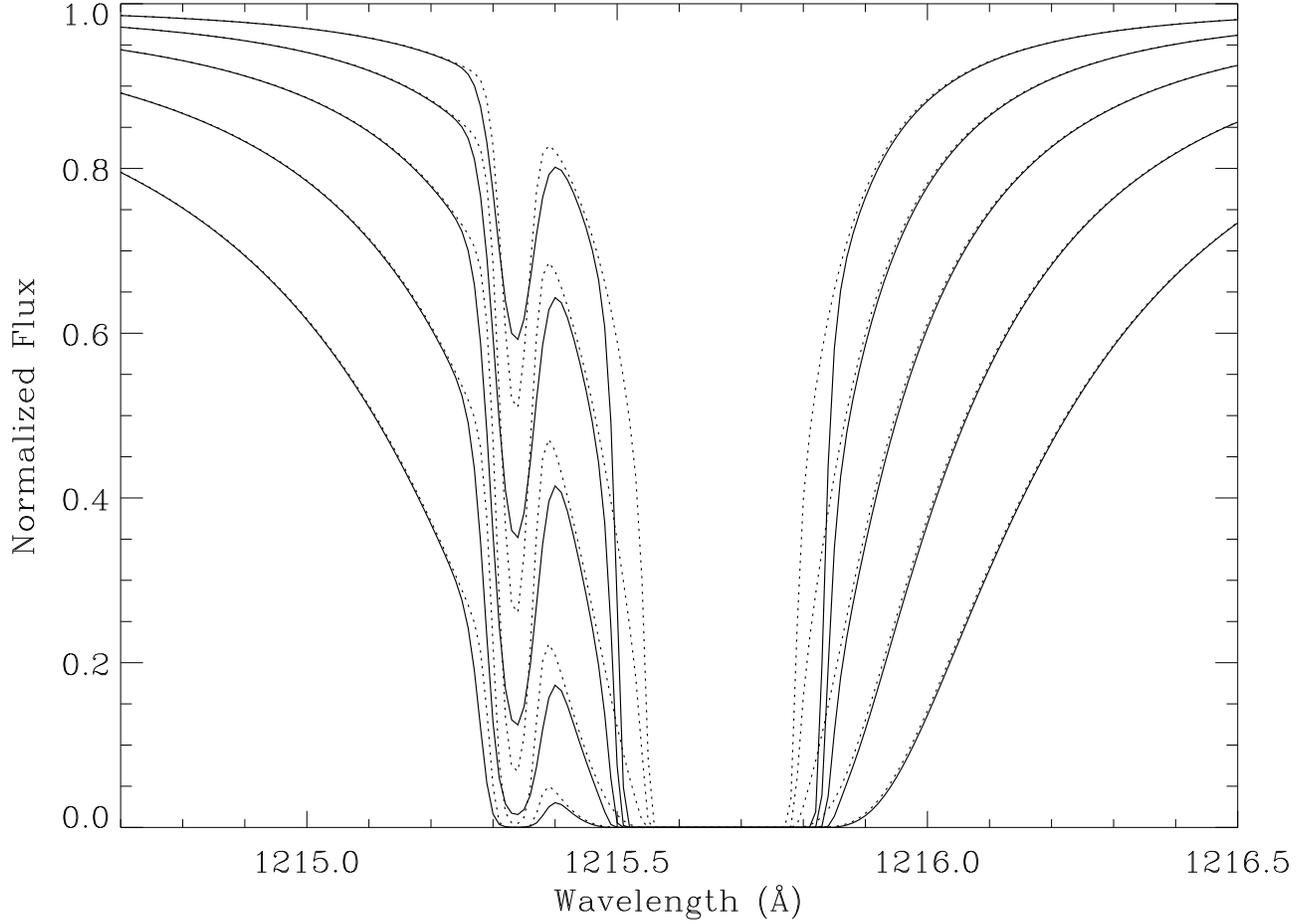}{3.5in}{90}{75}{75}{290}{0}
\caption{Absorption profiles for H~I+D~I Ly$\alpha$ absorption where
  the D~I and H~I parameters are related as described in the text,
  illustrating how absorption increases with increasing column density,
  where the profiles shown are for H~I column densities of
  $\log N({\rm H~I})=17.5$, 17.8, 18.1, 18.4, and 18.7.  The solid
  lines show the absorption for a Doppler parameter of
  $b({\rm H~I})=12.85$ km~s$^{-1}$ ($T=10,000$~K), while the dotted lines
  show the absorption for $b({\rm H~I})=9.09$ km~s$^{-1}$ ($T=5000$~K).}
\end{figure}

\begin{figure}
\plotfiddle{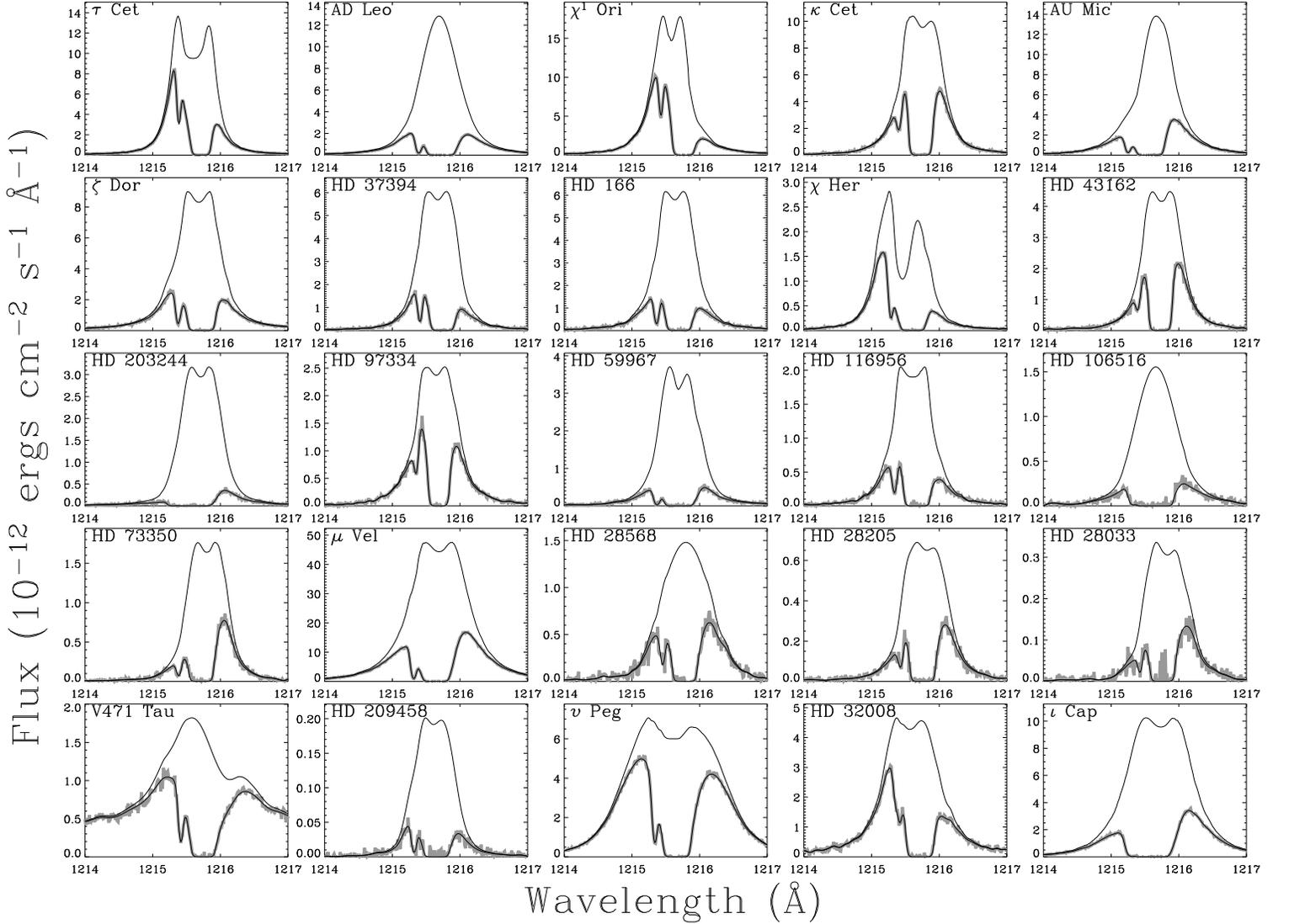}{5.0in}{90}{80}{80}{320}{20}
\caption{Our final best fits to the Ly$\alpha$ absorption spectra for the
  cases in which the ISM alone can account for all of the observed
  absorption.  In each panel, the upper solid line is the reconstructed
  stellar line profile and the lower solid line is the fitted absorption
  profile, which fits the data (shaded line).  The fit parameters are
  listed in Table~2, and integrated fluxes of the reconstructed stellar
  Ly$\alpha$ emission lines are listed in Table~3.}
\end{figure}

\begin{figure}
\plotfiddle{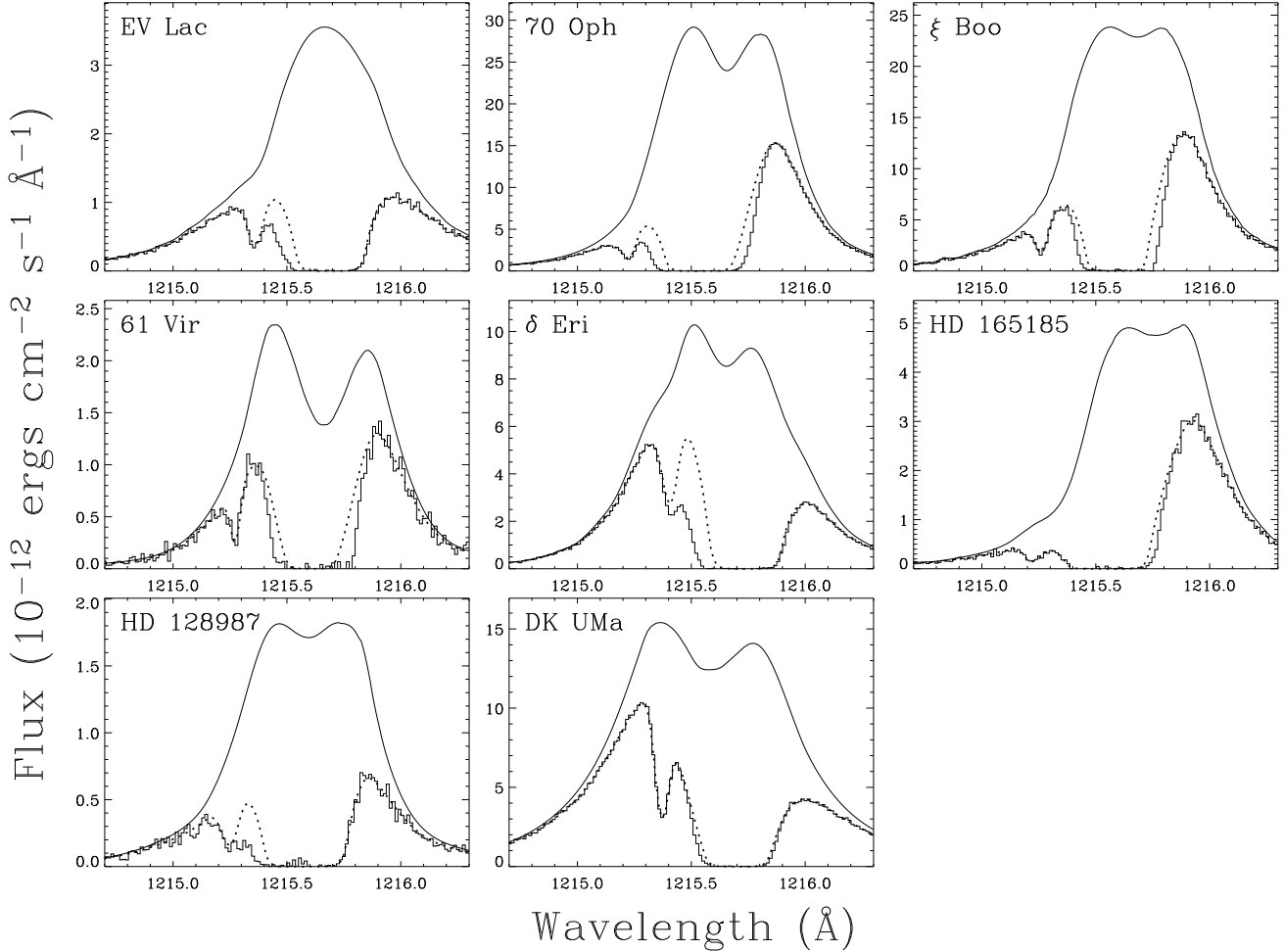}{3.5in}{90}{75}{75}{290}{0}
\caption{Our best fits to the Ly$\alpha$ absorption spectra for the cases
  in which HS/AS absorption components must be included in fits to the data,
  in addition to the ISM absorption.  Rather than show the total absorption
  implied by the fit, we show the absorption only from the ISM absorption
  components (dotted lines), the discrepancy with the data indicating the
  excess absorption that must be accounted for by the HS/AS components.
  Excess absorption on the blue side of the line is interpreted as
  astrospheric absorption, and excess absorption on the red side of the
  line is assumed to be heliospheric.  The fit parameters of these fits are
  listed in Table~2, and integrated fluxes of the reconstructed stellar
  Ly$\alpha$ emission lines are listed in Table~3.}
\end{figure}

\begin{figure}
\plotfiddle{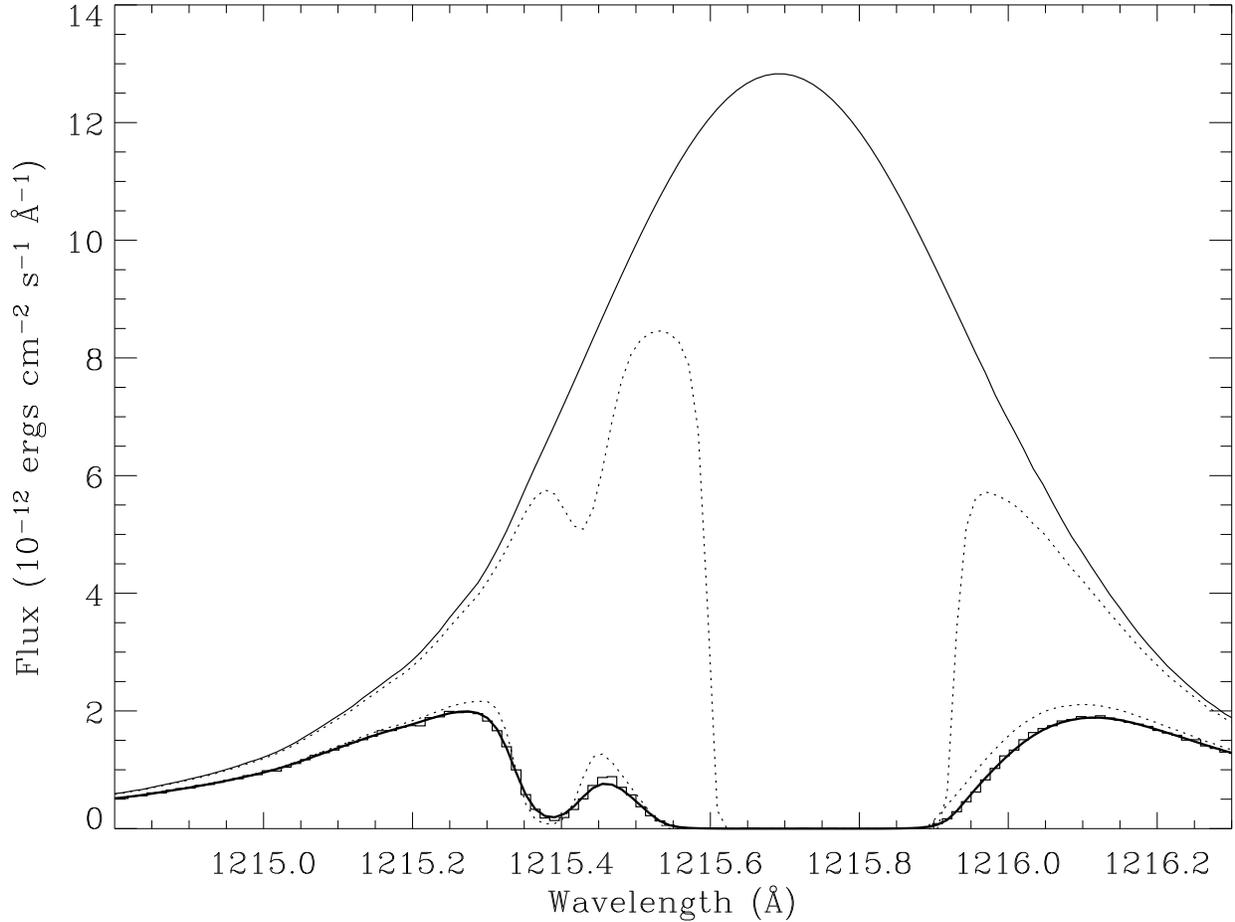}{3.5in}{90}{75}{75}{290}{0}
\caption{A two-component fit to the H~I+D~I Ly$\alpha$ absorption line
  seen towards AD~Leo.  The weaker component is assumed to be redshifted
  relative to the primary component by 12 km~s$^{-1}$ and is forced to
  have a column density 10 times lower than the primary component.  The
  Doppler parameters of the two components are forced to be
  equivalent, and in the fit they end up at
  $b({\rm H~I})=12.85$ km~s$^{-1}$.  This is significantly lower than
  the value found for the single component fit in Fig.~7 (see Table~2),
  illustrating how multiple ISM components can artificially increase
  measured Doppler parameters when assuming a single component.}
\end{figure}

\begin{figure}
\plotfiddle{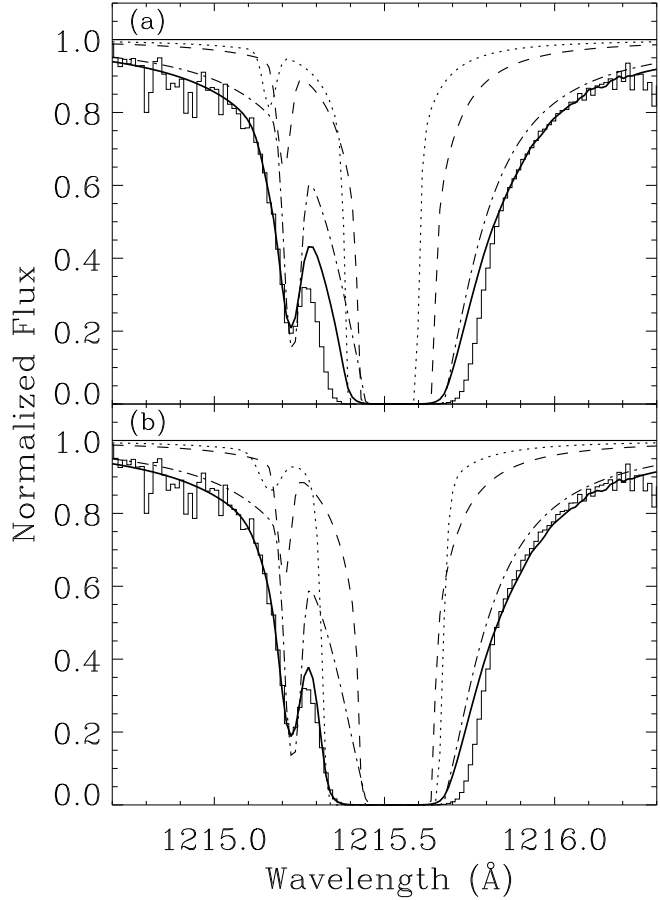}{3.5in}{90}{75}{75}{290}{0}
\caption{(a) A reproduction of the best 70~Oph fit from Fig.~7, but
  showing the individual ISM components as well as the total ISM
  absorption (thick solid line), where the dot-dashed, dashed, and dotted
  lines are absorption by ISM components 1, 2, and 3, respectively.
  The fit parameters are listed in Table~2.  The excess absorption
  seen on the red and blue sides of the H~I absorption is assumed to
  be heliospheric and astrospheric absorption, respectively.  (b) A fit
  to the 70~Oph data like that in (a), but with component~3 (dotted line)
  allowed to have a Doppler parameter large enough to account for the
  excess absorption on the blue side of the H~I absorption line.
  In this fit, the component~3 Doppler parameter is too large to be
  consistent with that measured for this component from Mg~II absorption,
  so we conclude that this fit is unreasonable (see text).}
\end{figure}

\begin{figure}
\plotfiddle{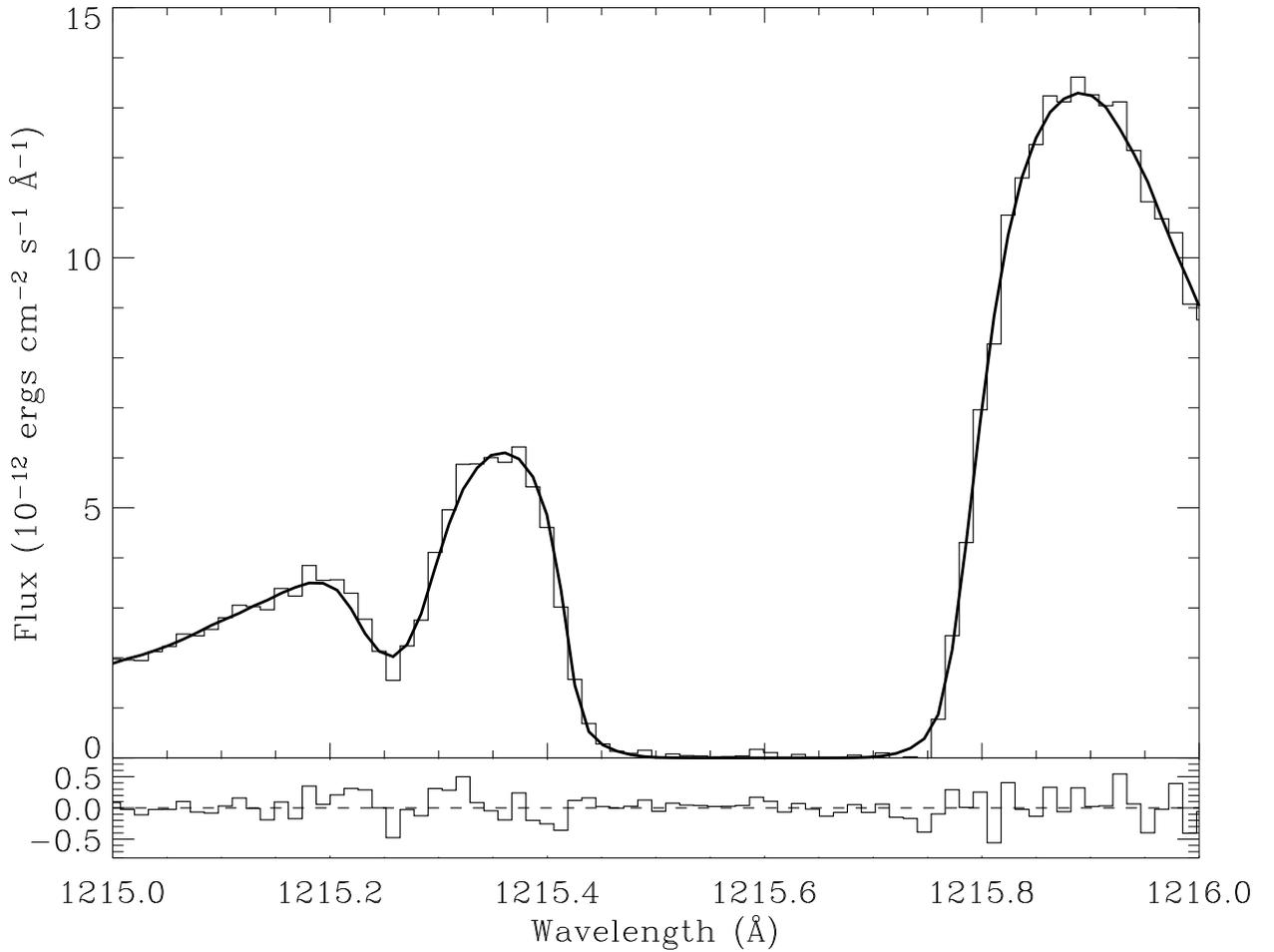}{3.5in}{90}{75}{75}{290}{0}
\caption{A fit to the Ly$\alpha$ absorption seen towards $\xi$~Boo~A, with
  residuals shown below the fit.
  The centroid of the HS/AS absorption is fixed such that it can
  only account for excess H~I absorption on the red side of the line
  (i.e., heliospheric absorption) rather than the excesses on both sides
  of the line suggested by the fit in Fig.~7.  This results in a poor fit to
  the D~I absorption at 1215.25~\AA.  The D~I fit is much too broad,
  demonstrating that the ISM H~I and D~I absorption must be narrower and the
  HS/AS component must therefore be allowed to contribute to the absorption
  on the blue side of the H~I line as well as the red side.  The
  implication is that astrospheric absorption is present towards $\xi$~Boo.}
\end{figure}

\begin{figure}
\plotfiddle{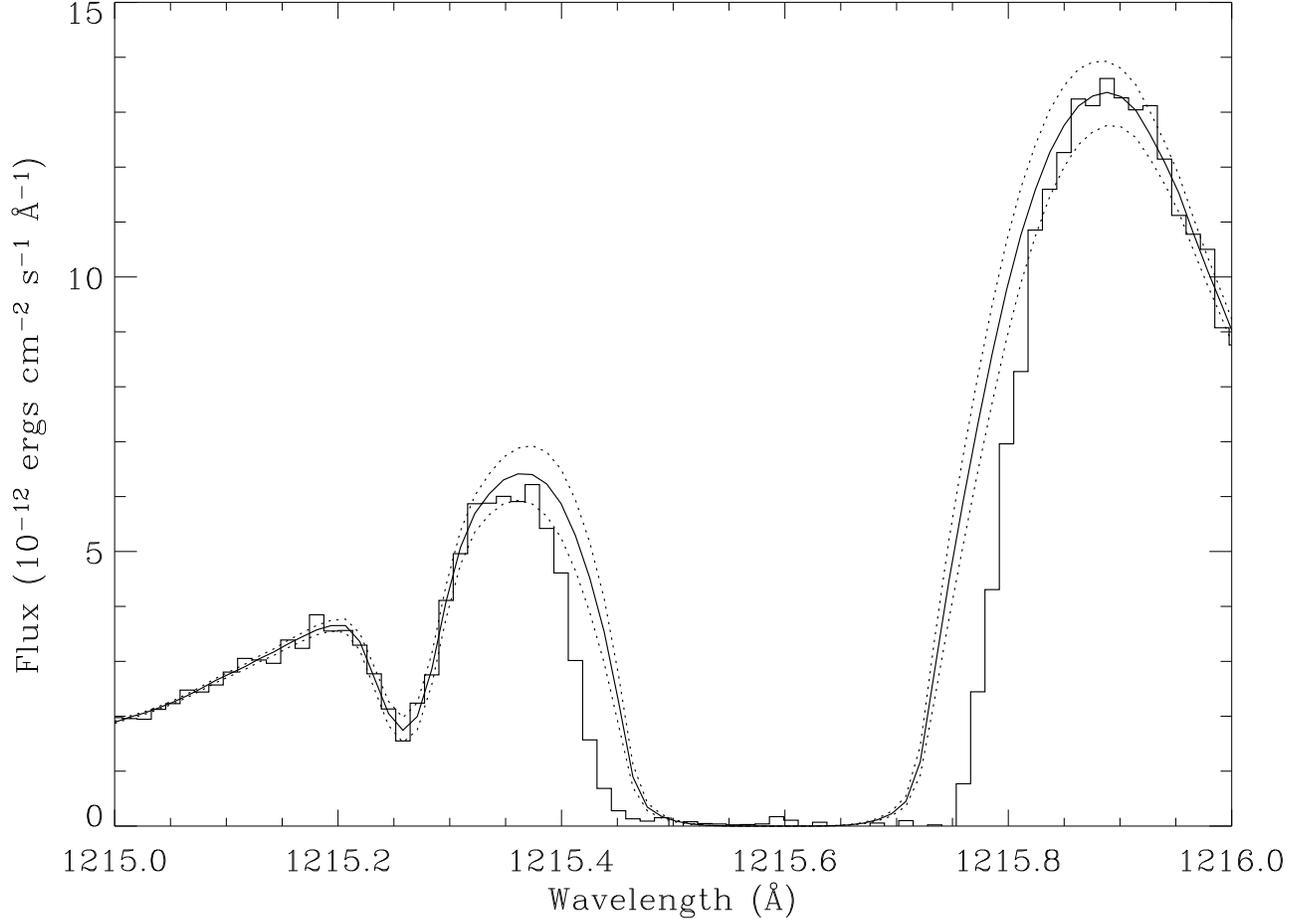}{3.5in}{90}{75}{75}{290}{0}
\caption{A reproduction of the best $\xi$~Boo fit from Fig.~7 (solid line),
  but also showing the absorption profile that results when the H~I
  column density is increased and decreased by 10\% (dotted lines).
  A 10\% increase in the column density clearly does not broaden the
  absorption enough to account for the excess H~I absorption that we
  interpret as heliospheric and astrospheric absorption.}
\end{figure}

\begin{figure}
\plotfiddle{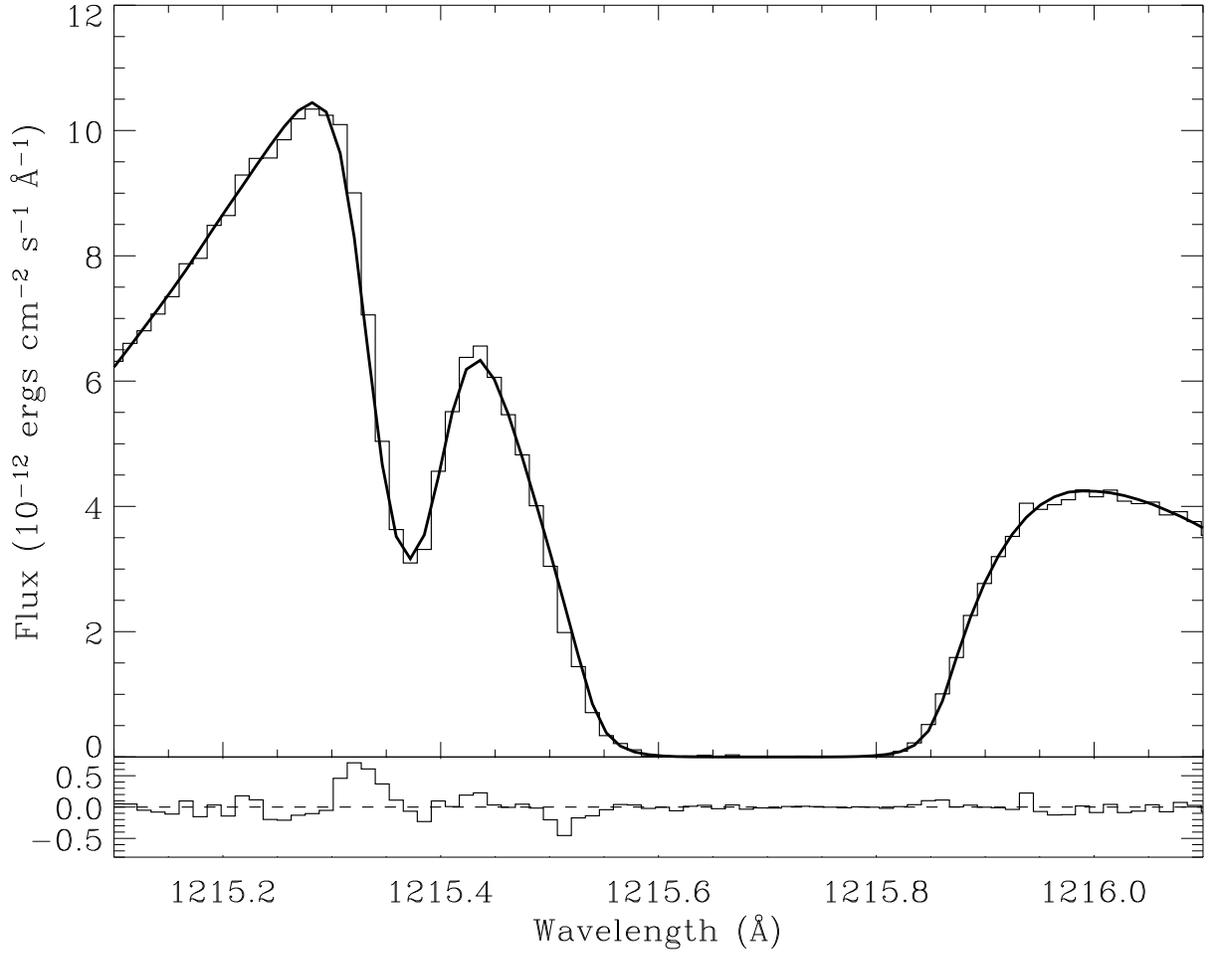}{3.5in}{90}{75}{75}{290}{0}
\caption{The best fit possible to the Ly$\alpha$ absorption observed
  towards DK~UMa assuming that the absorption is entirely from the ISM,
  with residuals shown below the fit.  The fit is poor along the blue sides
  of both the D~I and H~I lines near 1215.33~\AA\ and 1215.51~\AA,
  respectively.  These discrepancies can be fixed by the addition of an
  HS/AS absorption component, as shown in Fig.~7.  The discrepancies are
  somewhat subtle, but the S/N of the data is high enough for us to conclude
  that it is significant and indicative of at least a marginal detection of
  astrospheric absorption.}
\end{figure}

\begin{figure}
\plotfiddle{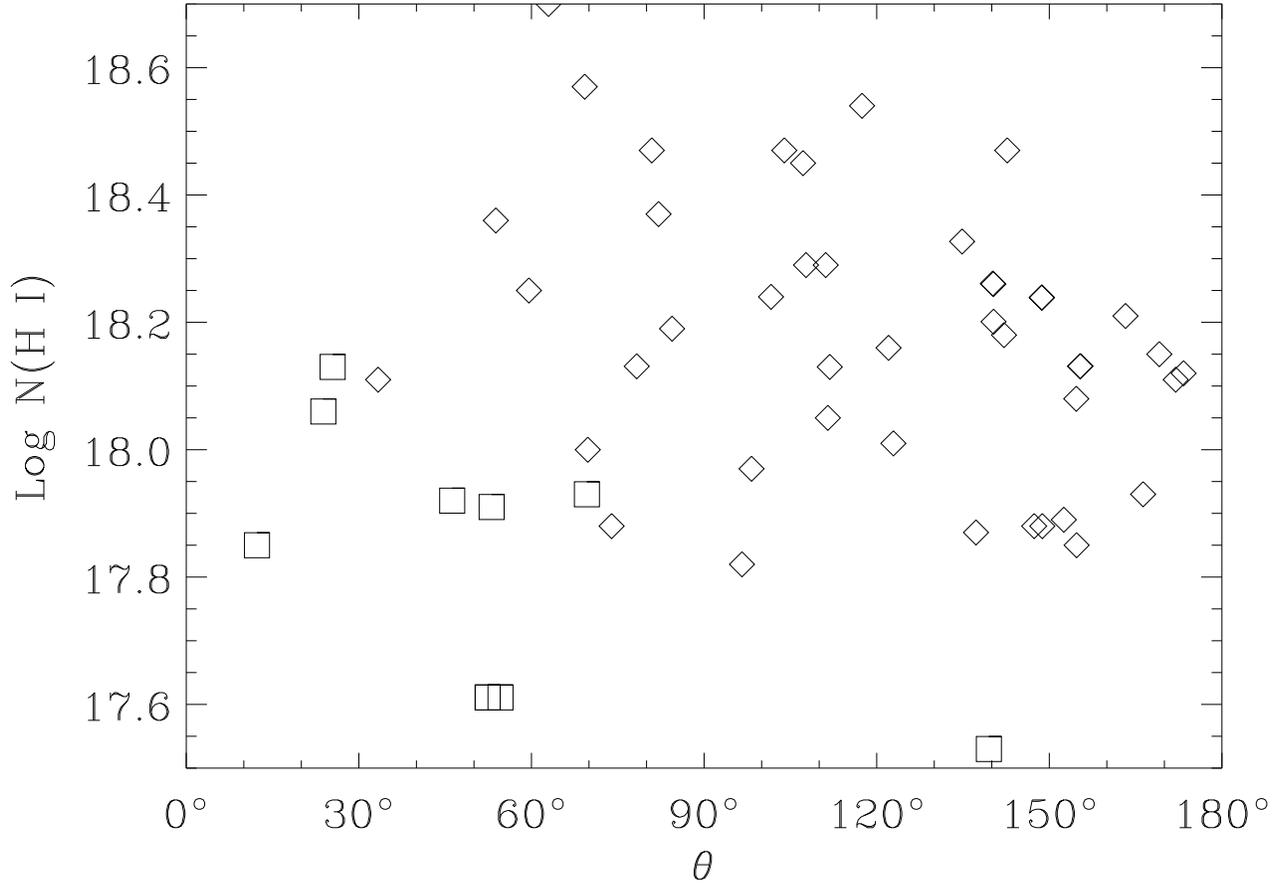}{3.5in}{90}{75}{75}{290}{0}
\caption{The ISM H~I column densities measured for all HST-observed lines
  of sight are plotted versus the angle of the lines of sight relative to
  the upwind direction of the ISM flow seen by the Sun.  The boxes and
  diamonds indicate lines of sight that yield detections and nondetections
  of heliospheric absorption, respectively.}
\end{figure}

\begin{figure}
\plotfiddle{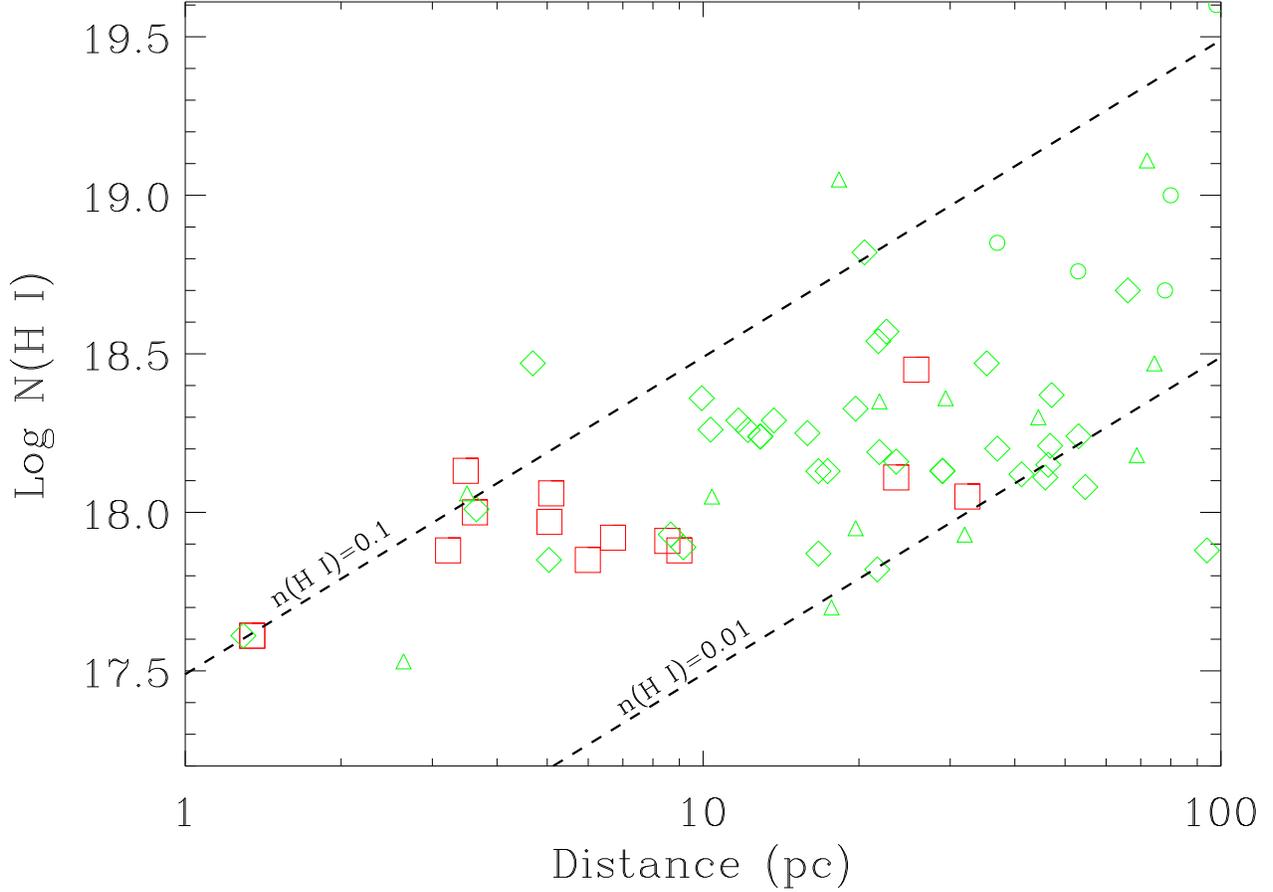}{3.5in}{90}{75}{75}{290}{0}
\caption{The ISM H~I column densities measured for all HST-observed lines
  of sight are plotted versus distance.  The red boxes are lines of sight
  that yield detections of astrospheric absorption, the diamonds are lines
  of sight that yield nondetections, and the triangles are lines of sight
  for which a valid search for astrospheric absorption is not possible
  (see text).  Finally, the circles are a few additional measurements that
  are from EUVE or {\em Copernicus} instead of HST.  The dashed lines are
  contours for line-of-sight average H~I densities of
  $n({\rm H~I})=0.1$ cm$^{-3}$ and $n({\rm H~I})=0.01$ cm$^{-3}$.}
\end{figure}

\begin{figure}
\plotfiddle{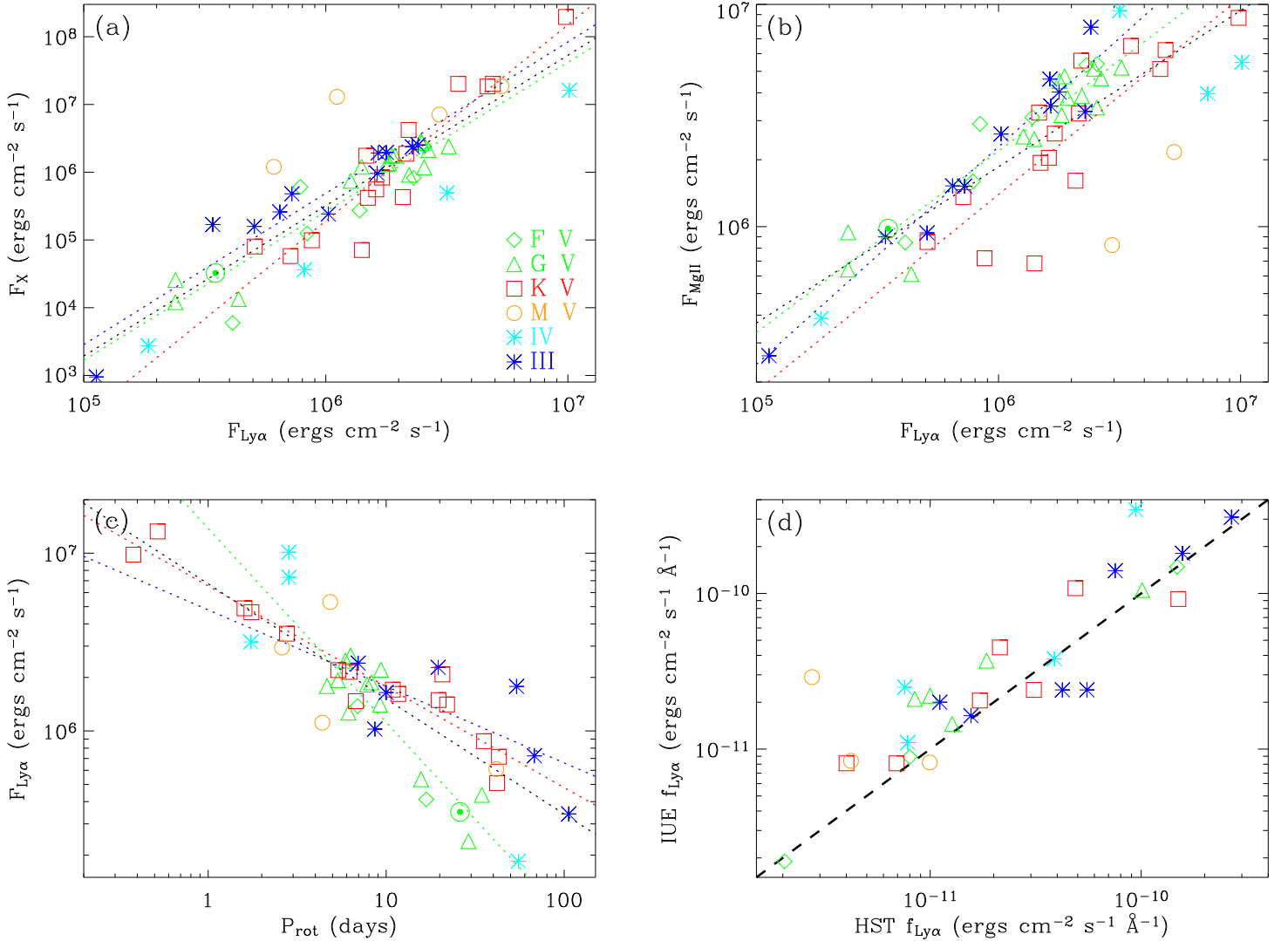}{5.0in}{90}{85}{85}{330}{-10}
\caption{(a) X-ray surface flux plotted versus Ly$\alpha$ surface flux
  for the stars listed in Table~3, with the addition of a point representing
  the Sun ($\odot$).  Different symbols are used for different
  classes of stars.  Dotted lines show various fits to the data, where the
  color of the line corresponds to the color of the data points being
  fitted.  The black dotted line is a fit to all the data combined.
  (b) Mg~II h \& k surface flux plotted versus Ly$\alpha$ surface flux,
  where the lines and symbols are as in (a).  (c) Ly$\alpha$ surface flux
  plotted versus rotation period, where the lines and symbols are as in (a).
  (d) Ly$\alpha$ fluxes measured by IUE from Landsman \& Simon (1993)
  plotted versus the HST measurements, where the symbols are as in (a).
  The dashed line is the line of equivalency.  Note that the parameters of
  the fits shown in (a)--(c) are listed in Table~4.}
\end{figure}

\end{document}